\documentclass{aa}  

\usepackage{graphicx}
\usepackage{txfonts}
\begin{document}

   \title{A refined analysis of the low-mass eclipsing binary system T-Cyg1-12664
   \thanks{Tables 1, 2, 3 and 10 are only available in electronic form
at the CDS via anonymous ftp to cdsarc.u-strasbg.fr (130.79.128.5)
or via http://cdsweb.u-strasbg.fr/cgi-bin/qcat?J/A+A/}}


   \author{Ram\'on Iglesias-Marzoa\inst{1,2}
          \and
          Mercedes L\'opez-Morales\inst{3}
          \and
          Mar\'ia J. Ar\'evalo\inst{1,4}
          \and
          Jeffrey L. Coughlin\inst{5}
          \and
          Carlos L\'azaro\inst{1,4}
          }

   \institute{Astrophysics Department, Universidad de La Laguna,
	      E-38205 La Laguna, Tenerife, Spain
	      \and
             Centro de Estudios de F\'isica del Cosmos de Arag\'on,
             Plaza San Juan 1, E-44001, Teruel, Spain\\
             \email{riglesias@cefca.es} 
             \and
             Harvard-Smithsonian Center for Astrophysics,
             60 Garden Street, Cambridge, MA 02138, USA
             \and
             Instituto de Astrof\'isica de Canarias,
             E-38200, La Laguna, Tenerife, Spain
             \and
             SETI Institute, 189 Bernardo Ave, Mountain View, CA 94043, USA
             }

   \date{Received September 15, 1996; accepted March 16, 1997}


  \abstract
   {The observational mass-radius relation of main sequence stars with masses between $\sim$ 0.3 and 1.0 $M_{\sun}$
   reveals deviations between the stellar radii predicted by models and the observed radii of stars in detached binaries.}
   {We generate an accurate physical model of the low-mass eclipsing binary T-Cyg1-12664
   in the Kepler mission field to measure the physical parameters of its components and to
   compare them with the prediction of theoretical stellar evolution models.}
   {We analyze the Kepler mission light curve of T-Cyg1-12664 to accurately measure the times and phases of the
   primary and secondary eclipse. In addition, we measure the rotational period of the primary component by analyzing
   the out-of-eclipse oscillations that are due to spots. We accurately constrain the effective temperature
   of the system using ground-based absolute photometry in B, V, $R_C$, and $I_C$. We also obtain and analyze
   $VR_C I_C$ differential light curves to measure the eccentricity and the orbital inclination of the system, and a
   precise $T_{\rm eff}$ ratio. From the joint analysis of new radial velocities and those in the literature
   we measure the individual masses of the stars. Finally, we use the PHOEBE code to generate a physical model
   of the system.}
   {T-Cyg1-12664 is a low eccentricity system, located $d$=360$\pm$22 pc away from us, with an orbital period of
   $P=4.1287955(4)$ days, and an orbital inclination $i$=86.969$\pm$0.056 degrees.
   It is composed of two very different stars with an active G6 primary with $T_{\rm eff1}$=5560$\pm$160 K,
   $M_1$=0.680$\pm$0.045 $M_\sun$, $R_1$=0.799$\pm$0.017 $R_\sun$, and a M3V secondary star
   with $T_{\rm eff2}$=3460$\pm$210 K, $M_2$=0.376$\pm$0.017 $M_\sun$, and $R_2$=0.3475$\pm$0.0081 $R_\sun$.}
   {The primary star is an oversized and spotted active star, hotter than the stars in its mass range.
   The secondary is a cool star near the mass boundary for fully convective stars
   ($M\sim0.35$ $M_\sun$), whose parameters appear to be in agreement with low-mass stellar model.}


   \maketitle
%

\section{Introduction}

T-Cyg1-12664 (KIC~10935310, 2MASS~J19513982+4819553) is a low-mass detached eclipsing binary (LMDEB) at
$\alpha$=19:51:39.824, $\delta$=+48:19:55.38 (J2000.0). It was first discovered in the T-Cyg1 field of
the Trans-Atlantic Exoplanet Survey \citep[TrES;][]{alonso2004}, and it is also in the Kepler mission
field of view. Because of its intermediate orbital period ($\sim$4.129 d) and low mass ratio
(q $\simeq$ 0.55), this system can serve as a benchmark for low-mass ($M$<1 $M_\sun$) stellar
evolution models that have problems with the radii predicted for the stars.

The discovery and first analysis of T-Cyg1-12664 was reported by \cite{devor2008} and \cite{devor2008thesis}
using TrES photometric data. They obtained a small number of radial velocity (RV)
observations, including only one RV measurement of the faint secondary star.
Their RV measurements show that the period initially identified in the TrES photometry (8.26 days)
was in fact of 4.129 days, and a secondary eclipse would be buried in the TrES data.
From that first analysis \cite{devor2008thesis} estimate masses of about 0.62 $M_\sun$ and
0.32 $M_\sun$ for the primary and secondary. \cite{cakirli2013} reanalyzed this system using
new RV measurements, the high precision Kepler mission light curves and $R$-band ground-based
photometry and obtained the following values for the masses and radii of the stars:
$M_1$ = 0.680$\pm$0.021 $M_\sun$, $R_1$ = 0.613$\pm$0.007 $R_\sun$;
and $M_2$ = 0.341$\pm$0.012 $M_\sun$, $R_2$ = 0.897$\pm$0.012 $R_\sun$.
In their study, while the radius of the primary is consistent with main sequence low-mass stellar models,
the radius of the secondary defies interpretation as a main-sequence star.

After its publication in \cite{devor2008}, we included T-Cyg1-12664 in  our observing program to
refine the physical parameters of several LMDEBs. Here we present a thorough analysis of the system
based on the Kepler photometry, new ground-based multiband photometry and RV measurements, and a
specific photometric calibration carried out to obtain unbiased colors and precise, photometry-based
effective temperatures.  The outline of this work is as follows.
In Sects.~\ref{sec:ligth-curve_observations} and \ref{sec:RVobservations} we describe the Kepler and
the ground-based RV and photometric observations. Section \ref{sec:analysis} describes a careful
analysis of the system, including specific sections for a consistent photometric calibration,
effective temperature, a careful treatment of the third light, and the orbital eccentricity.
Finally, the absolute parameters and the distance of the system are computed in
Sect.~\ref{sec:system_TCyg1}, with Sect.~\ref{subsec:comparison_with_models} devoted to comparing
our results to the stellar models and previous results.

\section{Light-curve observations}
\label{sec:ligth-curve_observations}

\subsection{Kepler observations}
\label{subsec:Kepler_LC}
\begin{figure}
\centering
\includegraphics[width=\hsize]{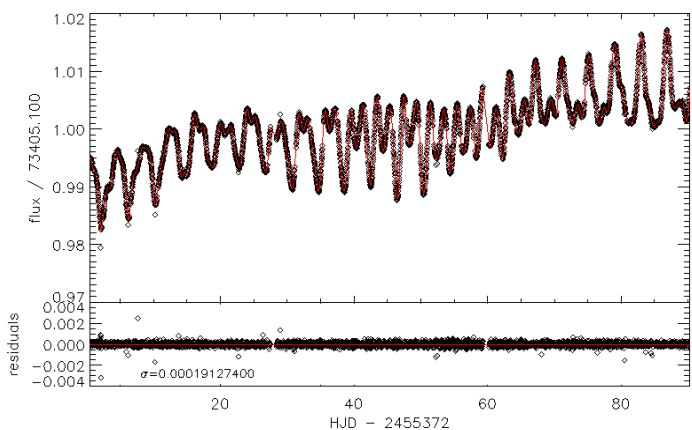}
  \caption{Fit of an order 300 Legendre polynomial to the Kepler data in quarter 6. The upper panel shows the
  out-of-eclipse light curve and the fit (red line). The lower panel shows the residuals of the fit.}
      \label{fig:detrendKepler_fit_and_residuals}
\end{figure}

\begin{figure}
\centering
\includegraphics[width=\hsize]{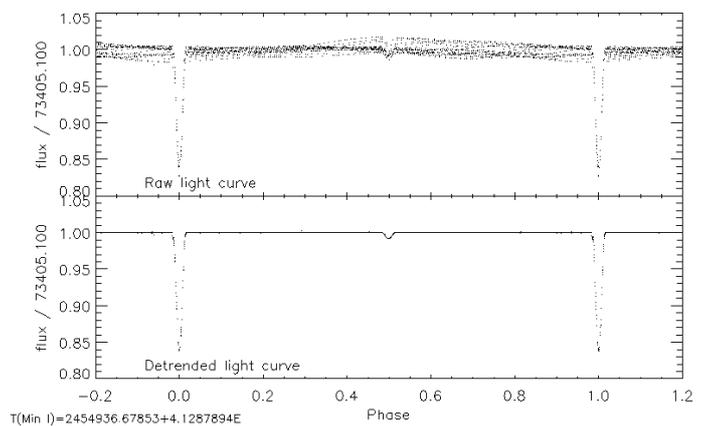}
  \caption{Effect of the detrending over the phased light curve of Q6 quarter. Upper panel shows the
  raw light curve and   lower panel shows the light curve after applying the Legendre polynomial
  fitted in Fig.~\ref{fig:detrendKepler_fit_and_residuals}.}
      \label{fig:detrendKepler_light_curve}
\end{figure}

T-Cyg1-12664 was observed by the Kepler spacecraft in Long Cadence mode from Q0 through Q17
(JD 2454953.54 -- JD 2456424.00). The observations were reduced using the Kepler mission pre-search
data conditioning (PDC) pipeline \citep{jenkins2010}, which produces high quality light curves like the ones
illustrated in Figs.~\ref{fig:detrendKepler_fit_and_residuals} and \ref{fig:detrendKepler_light_curve} for Q6.
The instrumental systematics and spot-related activity in this star are readily visible in the Kepler
light curves with out-of-eclipse variations deeper than the secondary eclipse.

We detrended the light curves from each quarter following a procedure similar to \cite{slawson2011}, by fitting
Legendre polynomials to the out-of-eclipse data. The result for one quarter (Q6) is shown in
Fig.~\ref{fig:detrendKepler_light_curve}. For T-Cyg1-12664 we had ephemeris information from preliminary works,
so we could exclude the time intervals affected by eclipses, and it was not necessary to apply the iterative processes
described in \cite{slawson2011}. Because of the effects of the spacecraft safe-modes and triggers, some
quarters were divided in smaller intervals, but this detrending process was applied
to all the Kepler mission photometry encompassing an interval of 1470 days. Typical Legendre polynomials orders were
between 30 and 300, depending on the complexity of the variations and duration of the light-curves intervals.
The result of this detrending is illustrated in the bottom panel of Fig.~\ref{fig:detrendKepler_light_curve}.
This detrended light curve, provided in Table~\ref{table:detrended}, was subsequently analyzed together with
the ground-based photometry described in Sect.~\ref{subsec:ground-based_phot}.

We note that this detrending process has the disadvantage of potentially removing information from
the light curve, such as heating effects and proximity effects that are due to the shape of the stars.
However, as we demonstrate in subsequent sections, these effects are negligible
in T-Cyg1-12664, because of the large separation between the stars.

\onltab{
\begin{table*}
\caption{Detrended Kepler light curve for T-Cyg1-12664. Full version of this table is avaliable in
electronic form at the CDS.}
\label{table:detrended}
\centering
\begin{tabular}{cc}
\hline  \hline
BJD             &Normalized flux\\
day             &--\\
\hline
2454954.43757	&1.0000260	\\
2454954.45800	&0.99973979	\\
2454954.47843	&1.0001384	\\
...&\\
\hline
\end{tabular}
\end{table*}
}

\subsection{Ground-based observations}
\label{subsec:ground-based_phot}

We observed T-Cyg1-12664 with the CAMELOT camera at the IAC80 telescope in Tenerife, Spain, in $VR_CI_C$ bands,
over 14 nights, between 2009-04-04 and 2010-10-01 UT (JD 2454936.6 -- JD 2455471.5), when eclipses occurred.
We also requested a routinary observation program at the same telescope to sample out-of-eclipse phases.
CAMELOT is equipped with a back-illuminated E2V 2048x2048 sensor with squared 13.5 $\mu m$ pixels,
which results in a plate scale of 0.''304/pix and a 10.'4x10.'4 field of view. 

Although CAMELOT has two readout channels, all the observations were made using one channel to avoid systematic
effects between the object and the comparison stars. Exposure times ranged from 100 s to 180 s in $V$ band,
65 s to 95 s in $R_C$ band, and 60 s to 90 s in $I_C$ band, depending on the seeing and the atmospheric
transparency. Each night, we monitored T-Cyg1-12664 continuously over an altitude limit of 30 degrees
(airmass $\leq$ 2) above the horizon. We also acquired bias frames and dome flats in all the filters at the
beginning of each night.

We used standard reduction techniques, including bias substraction, flat-field correction, and
alignment of all images to a common reference system, to process the frames.
We analyzed the corrected frames using the \verb;phot; aperture photometry package in IRAF and a
custom differential photometry pipeline designed to deal with large sets of frames.
The pipeline first performs aperture photometry for a set of preselected comparison stars and
the target, for a range of aperture sizes. Using the flux within each aperture, the pipeline
computes the signal-to-noise ratio (S/N) of the target using the CCD equation
\citep[][eq.~25]{merline1995}, and selects the aperture radius that produces the maximum S/N.
This aperture is then used to extract the flux of all the stars in that frame. This procedure accounts
for seeing and transparency variations along the night. The CCD equation was also used to compute
the formal errors in the photometry. We imposed a limit of 9 pixels to the maximum aperture radius to avoid
contamination by nearby stars as is the case for T-Cyg1-12664 (see Sect.~\ref{subsec:thirdlight}).
The selected comparison stars proved to be stable during all the observing runs.
We generated $V$, $R_C$, and $I_C$ differential light curves of the target dividing its flux by the
combined flux of all  the comparison stars. Owing to the small
field of view of the telescope no differential extinction effects were taken into account.
Our $VR_CI_C$ photometry (given in Table~\ref{table:IAC80phot}) includes seven primary and four
secondary eclipses which, when added to those from the Kepler, \cite{cakirli2013} and TrES photometry,
enable us to improve the period of the system and search for period variations.

\onltab{
\begin{table*}
\caption{IAC80 differential photometry for T-Cyg1-12664. Full version of this table is avaliable in
electronic form at the CDS.}
\label{table:IAC80phot}
\centering
\begin{tabular}{ccccc}
\hline  \hline
BJD                 &$\Delta$m&Error&Airmass&Filter\\
day                 &mag    &mag    &--     &--\\
\hline
2455378.404710830	&1.603	&0.001	&1.825	&V	\\
2455378.407322901	&1.671	&0.001	&1.796	&Rc	\\
2455378.409550562	&1.767	&0.001	&1.771	&Ic	\\
...&\\
\hline
\end{tabular}
\end{table*}
}

\section{Radial velocity observations}
\label{sec:RVobservations}

We observed T-Cyg1-12664 as part of a larger program to produce accurate radial velocity curves
of low-mass binaries \cite[see][]{coughlin2012}. We collected radial velocity observations over
two seasons (June -- December 2010 and May -- September 2011), with the Dual Imaging Spectrograph (DIS)
on the 3.5-m at Apache Point Observatory and the R-C Spectrograph on the 4-m at Kitt Peak National
Observatory (KPNO). With DIS we observed using its red channel with the R1200 grating, which gives a
resolution of 0.58~\AA~per pixel, or R~$\approx$~10,000. The wavelength range was set to
$\sim$5900 - 7100~\AA~for the first observing season, and $\sim$5700 - 6800~\AA~for the second.
For the R-C Spectrograph, we used the KPC-24 grating in second order, resulting in a resolution
of 0.53~\AA~per pixel, or R~$\approx$~10,000, with the wavelength range set to $\sim$5700 - 6750~\AA.
We collected a total of 31 individual spectra for T-Cyg1-12664, in addition to bias, dark, and
flatfield calibration frames. We also collected HeNeAr lamp frames before or after each target
observation to use as wavelength calibrations.

All raw frames were bias, dark, and flat-field corrected. For the DIS observations,
column 1023 is a dead column, and thus we replaced those values by a linear interpolation
between the two neighboring columns. We also removed cosmic rays from all frames using a Laplacian
edge algorithm implemented in the \verb|lacos_spec| IRAF package \cite[see][]{vandokkum2001}.
Finally, we extracted one-dimensional spectra from each image, including sky background substraction,
using the IRAF package \verb|apall|. After wavelength calibration, all science spectra were flattened
and normalized by fitting a 20-piece cubic spline fit over ten iterations, during which points
3$\sigma$ above the fit or 1.5$\sigma$ below the fit were rejected. For each spectrum, we calculated
the barycentric Julian date in terrestrial time, BJD(TT), and corrected them to the reference frame
of the solar system barycenter using the IRAF task \verb|bcvcorr|.

For all binaries we attempted to extract the radial velocity of each star in each frame, $V_{1,j}$ and
$V_{2,j}$, where $1$ and $2$ denote the primary and secondary stars, and $j$ is the frame number,
using a new exploratory method: instead of using cross-correlations, we directly fit reference spectra
to each observed spectrum via a traditional standard deviation minimization, which is equivalent
to minimizing $\chi^{2}$, assuming all observed points have equal errors.
Specifically, for each observed spectrum, we fit for the velocity of each reference spectra, as well as for their
luminosity ratio, $\alpha$ = $L_1 / L_2$, for a total of three free parameters. During the fitting, the original
observed spectrum is never changed, and thus it can have any arbitrary number and distribution of wavelength and
flux value pairs. Given a total of $M$ observed spectra, each with $N_{j}$ points, taken at
times $t_{j}$, we want to find the values of $V_{1,j}$, $V_{2,j}$, and $\alpha$ at each $t_{j}$, that minimize the function
\begin{equation}
\label{rmseq}
  \sum_{j=1}^{M}{\sum_{i=1}^{N_{j}}{\left(F_{0,i} - \frac{\alpha\cdot F_{1}\left(\lambda_{0,i}\cdot(1-\frac{V_{1,j}}{c})\right) + F_{2}\left(\lambda_{0,i}\cdot(1-\frac{V_{2,j}}{c})\right)}{\alpha+1}\right)^{2}}} ,
\end{equation}

\noindent where $F_{0,i}$ is the flux of an observed spectrum at a given wavelength,
$\lambda_{0,i}$, $F_{1}$($\lambda$), and $F_{2}$($\lambda$) are the fluxes of reference
stars 1 and 2 at a given wavelength, and $c$ is the speed of light.
We determine $F_{1}$($\lambda$) and $F_{2}$($\lambda$) by cubic splines interpolation of the
original reference spectra.

For reference spectra, we used the normalized (flattened) synthetic spectra from \citet{munari2005}.
The \citet{munari2005} grid we used covers models with 3500 $\geq$ $T_{\rm eff}$ $\geq$ 10,000 in
steps of 250 K, 0.0 $\geq$ $\log{g}$ $\geq$ 5.0 in steps of 0.5 dex, -2.5 $\geq$ [M/H] $\geq$ 0.5 in
steps of 0.5 dex, and 0 $\geq$ $V_{rot}$ $\geq$ 100 km/s in steps of 10 km/s. To ensure that we
found the global minimum in both selected reference spectra and their velocities, we performed a global
grid search looping over all possible combinations of $\alpha$, $V_{1,j}$, $V_{2,j}$, $T_{1}$, $T_{2}$,
using a common [M/H] for both reference spectra, and $\log{g}$ and $V_{rot}$ for each reference star.
We compared the radial velocities found by this method with those obtained using the well-tested
TODCOR code \citep{zucker1994}, and find consistent results for all the binaries in our program.

The errors in the derived radial velocities were computed using a bootstrapping re-sampling method iterated
over 10000 times. The adopted errors correspond to the 1$\sigma$ confidence interval of the distribution of
bootstrapping solutions. To estimate the errors in the derived temperatures, we used the spectroscopic
quality-of-fit parameter \citep{lopez-morales2008,behr2003}, defined as
\begin{equation}
  z = \sqrt{N}\left(\frac{rms^{2}}{rms^{2}_{min}} - 1\right) , 
\end{equation}

\noindent where N is the number of data points, $rms^{2}$ is the standard deviation of the fit under consideration,
and $rms^{2}_{min}$ is the best fit found. The $z$ parameter is similar to a reduced $\chi^{2}$ in the absence of
known errors on the individual points. By definition $z$ = 0 at the best-fit, and the 1$\sigma$ confidence
interval corresponds to  $z$ = 1.

In the particular case of T-Cyg1-12664, the spectrum of the secondary is too weak, and we were unable to detect any
significant radial velocity measurements for the secondary star over any reasonable parameter ranges.
Thus, we treated the system as a single-lined binary by fixing the value of $\alpha$ to 9 999 and only fitting
for the parameters of the primary star $V_{1,j}$, $T_{\rm eff1}$, [M/H], $\log{g}$ and $V_{rot}$.
The resultant radial velocities are listed in Table~\ref{tab:RV1}. In addition, the best fit of the primary's spectra
to the grid of synthetic spectra gives a $T_{\rm eff}$=5750$\pm$250 K,
$\log g$=4.5$\pm$0.5, $[M/H]=$-0.5$\pm$0.5 and $v_r \sin i$=40$\pm$10 km s$^{-1}$. 
This $T_{\rm eff}$ is in agreement with the one obtained in Sect.~\ref{subsec:Teff} using photometric colors.
The metallicity appears to be lower than the typical for stars near the Galactic plane (see Section 6.2),
which would have near-solar metallicity. We attribute this result to the rather coarse sampling in metallicity
of our models, and therefore adopt solar metallicity for this system in the remaining analysis.

We note that, although this $\chi^2$ minimization technique yields values of the radial velocities
and other stellar parameters consistent with those obtained via other methods in the case of T-Cyg1-12664,
the technique can be prone to systematics introduced by the presence of correlated noise or not well
modeled spectral instrumental profiles, especially in the case of low S/N spectra.
It is therefore advisable to have results checked against other methods like TODCOR.

\onltab{
\begin{table*}
\caption{Measured RV for the primary component in the T-Cyg1-12664 system. Full version of this table is avaliable in
electronic form at the CDS.}
\label{tab:RV1}
\centering
\begin{tabular}{ccc}
\hline
\hline
BJD	            &RV1	&error\\
day             &km s$^{-1}$&km s$^{-1}$\\
\hline
2455426.823697	&49.82	&7.61	\\
2455447.710532	&60.73	&4.36	\\
2455450.826485	&7.83	&10.32	\\
...&\\
\hline
\end{tabular}
\end{table*}
}

\section{Analysis of the system}
\label{sec:analysis}

\subsection{Photometric calibration and colors}
\label{subsec:photcal_colors}

\begin{table}
\caption{\label{tab:catalog_phot} Photometry in several bands for T-Cyg1-12664
obtained from photometric catalogs.}
\centering
\begin{tabular}{ccl}
\hline
\hline
Bandpass&Magnitude		&Source\\
\hline
$V$	&13.520$\pm$0.190	&GSC\tablefootmark{a}\\
\hline
$B$	&14.040$\pm$0.416	&GSC2.3.2\tablefootmark{a}\\
$V$	&13.108$\pm$0.297	&GSC2.3.2\tablefootmark{a}\\
\hline
$B$	&13.530			&NOMAD\tablefootmark{b,c}\\
$V$	&13.120			&NOMAD\tablefootmark{b,c}\\
$R$	&13.500			&NOMAD\tablefootmark{b}\\
\hline
$B1$	&14.460			&USNO-B1\tablefootmark{d}\\
$B2$	&14.020			&USNO-B1\tablefootmark{d}\\
$R1$	&13.130			&USNO-B1\tablefootmark{d}\\
$R2$	&13.500			&USNO-B1\tablefootmark{d}\\
$I2$	&13.210			&USNO-B1\tablefootmark{d}\\
\hline
$r$	&12.400			&USNO-A2\tablefootmark{e}\\
$b$	&13.100			&USNO-A2\tablefootmark{e}\\
\hline
$B$	&13.530			&UCAC3\tablefootmark{e,f}\\
$R2$	&12.512			&UCAC3\tablefootmark{e,f}\\
$I$	&11.858			&UCAC3\tablefootmark{e,f}\\
\hline
$U$	&13.905$\pm$0.021	&\cite{everett2012}\\
$B$	&13.864$\pm$0.024	&\cite{everett2012}\\
$V$	&13.280$\pm$0.018	&\cite{everett2012}\\
\hline
$g'$	&13.541			&Sloan Digital Sky Survey\\
$r'$	&13.052			&Sloan Digital Sky Survey\\
$i'$	&12.911			&Sloan Digital Sky Survey\\
$z'$	&12.843			&Sloan Digital Sky Survey\\
\hline
$D51$	&13.327			&Kepler mission\tablefootmark{g}\\
$Kepler$&13.100			&Kepler mission\\
\hline
$J$	&11.911$\pm$0.024	&2MASS\tablefootmark{h}\\
$H$	&11.582$\pm$0.015	&2MASS\tablefootmark{h}\\
$K_s$	&11.529$\pm$0.021	&2MASS\tablefootmark{h}\\
\hline
\end{tabular}
\tablefoot{\\
\tablefoottext{a}{Identifier: GSC 0356101711.}\\
\tablefoottext{b}{Identifier: 1383-0349082.}\\
\tablefoottext{c}{Mag from YB2.}\\
\tablefoottext{d}{Identifier: 1383-0345184.}\\
\tablefoottext{e}{Identifier: U1350 11189502.}\\
\tablefoottext{f}{Mag from SuperCosmos.}\\
\tablefoottext{g}{Band centered in 510 nm, see \cite{brown2011}.}\\
\tablefoottext{h}{Identifier: 2MASS J19513982+4819553.}
}
\end{table}

In Table~\ref{tab:catalog_phot}, we compile all the available photometry for T-Cyg1-12664 in public catalogs.
We found inconsistencies between some of the passband magnitudes reported by different catalogs, e.g.
the B and V magnitudes reported by the GSC and NOMAD catalogs disagree by 0.4 -- 0.5 mags.
The differences are too large to be explained by stellar activity, or even accounting for the depth of
the eclipses ($\sim$0.20 in the Kepler band). We do not have an explanation for these discrepancies but,
concerned about introducing large errors in the determination of parameters for this system, we
proceeded to derive new, consistent colors.

We performed photometric calibrations of several objects, including T-Cyg1-12664, over two photometric
nights in July and August 2012, during out-of-eclipse phases. We used the Tr\"omso CCD Photometer in
full frame mode \citep[TCP, ][]{ostensen2000thesis,ostensen2000}, at the IAC80 telescope
(CAMELOT was not available at the time of these observations). The TCP is built around a Texas
Instruments 1024x1024 CCD sensor and produces a 9.'6 field of view, with a plate scale of 0.''537 pixel$^{-1}$,
at the IAC80's focal plane. 

We observed several Landolt standard fields \citep{landolt2009}, covering objects with a wide range of colors and
airmasses, and using Johnson-Cousins $B$, $V$, $R_C$, and $I_C$ filters. A log of the observations is shown
in Table~\ref{tab:LandoltFields}. We measured the flux of the standard stars via standard aperture photometry,
using the IRAF package \verb|apphot|, and adopting an aperture of 14'' (a 13 pixel radius in the TCP images),
to match the Landolt aperture widths.

\begin{table}
{\footnotesize
\caption{Observed Landolt fields. The number of measurements is the global count
for different air masses.}
\label{tab:LandoltFields}      
\centering                          
\begin{tabular}{c c c c c}        
\hline\hline                 
Night		&Field		&N$_{stars}$	&N$_{meas}$		&Airmass range\\
\hline
2012-07-26	&SA110SF3	&8		&19			&1.130-2.519\\
		&PG1633+099	&8		&9\tablefootmark{a}	&1.055-2.590\\
\hline
2012-08-18	&SA110SF3	&8		&4\tablefootmark{b}	&1.158-2.435\\
		&PG0231+051	&6		&9\tablefootmark{c}	&1.086-1.690\\
		&PG1657+078	&6		&6			&1.076 - 2.542\\
\hline
\end{tabular}
\tablefoottext{a}{8 measurements for the $B$ and $V$ filters.}
\tablefoottext{b}{6 measurements for the $I_C$ filter.}
\tablefoottext{c}{10 measurements for the $V$ filter.}
}
\end{table}

The extinction correction and transformation to the standard photometric system was performed
simultaneously solving the set of equations
{\footnotesize
\begin{equation}\label{eq:fotcal2}
\begin{alignedat}{7}
&m_B    &{}={}& B   &{}+{}& b_0 &{}+{}& b_1 X_B     &{}+{}& b_2 (B-V)   &{}+{}& b_3 X_B (B-V)       &{}+{}& b_4 (B-V)^2\\
&m_V    &{}={}& V   &{}+{}& v_0 &{}+{}& v_1 X_V     &{}+{}& v_2 (B-V)   &{}+{}& v_3 X_V (B-V)       &{}+{}& v_4 (B-V)^2\\
&m_{R_C}&{}={}& R_C &{}+{}& r_0 &{}+{}& r_1 X_{R_C} &{}+{}& r_2 (V-R_C) &{}+{}& r_3 X_{R_C} (V-R_C) &{}+{}& r_4 (V-R_C)^2\\
&m_{I_C}&{}={}& I_C &{}+{}& i_0 &{}+{}& i_1 X_{I_C} &{}+{}& i_2 (V-R_C) &{}+{}& i_3 X_{I_C} (V-R_C) &{}+{}& i_4 (V-R_C)^2\\
\end{alignedat}
\end{equation}
}
\noindent where the instrumental magnitudes appear on the left-hand side of the equations, the absolute
magnitudes are denoted as $B$, $V$, $R_C$, and $I_C$, and the airmass values as $X$.
The equations were solved using least-square techniques, and the resulting coefficients for each night
are summarized in Table~\ref{tab:stdcoefs}.  The absolute magnitudes obtained for T-Cyg1-12664 as the
average of the results for each night are summarized in Table~\ref{tab:calibrated_phot}. In that same
table, we also list color indexes, including the near-infrared $JHK_S$ bands
from 2MASS \citep{skrutskie2006}.

\begin{table}
\caption{Standard transformation coeficients from Eq.~\ref{eq:fotcal2} for the two nights.
The uncertainties are shown in parentheses affecting the last two digits.}             
\label{tab:stdcoefs}      
\centering                          
\begin{tabular}{c | c c}        
\hline\hline                 
Parameter	&2012-07-26	&2012-08-18\\    
\hline                        
$b_0$		&2.7185(30)	&2.7947(65)\\
$b_1$		&0.2262(18)	&0.2250(38)\\
$b_2$		&0.0101(39)	&0.0430(90)\\
$b_3$		&-0.0204(22)	&-0.0342(49)\\
$b_4$		&-0.0038(13)	&-0.0186(27)\\
\hline
$v_0$		&2.7752(25)	&2.8607(44)\\
$v_1$		&0.1184(13)	&0.1170(24)\\
$v_2$		&0.0436(30)	&0.0575(56)\\
$v_3$		&0.0025(13)	&-0.0082(27)\\
$v_4$		&0.01621(82)	&0.0163(15)\\
\hline
$r_0$		&2.8779(29)	&2.9391(46)\\
$r_1$		&0.0877(13)	&0.0905(24)\\
$r_2$		&-0.0662(61)	&0.029(11)\\
$r_3$		&-0.0096(21)	&-0.0208(40)\\
$r_4$		&0.0940(29)	&0.0501(56)\\
\hline
$i_0$		&3.2383(37)	&3.3249(52)\\
$i_1$		&0.0426(17)	&0.0482(29)\\
$i_2$		&-0.2608(78)	&-0.209(12)\\
$i_3$		&0.0024(26)	&-0.0326(45)\\
$i_4$		&0.0717(39)	&0.0865(73)\\
\hline                                   
\end{tabular}
\end{table}

\begin{table}
\caption{Calibrated magnitudes and color indices for T-Cyg1
obtained from our photometric calibration. The infrared magnitudes
are taken from 2MASS (see Table~\ref{tab:catalog_phot}).}
\label{tab:calibrated_phot}      
\centering                          
\begin{tabular}{l c}        
\hline\hline                 
Band		&Adopted value (mag)\\    
\hline                        
$B$		&13.937$\pm$0.002\\
$V$		&13.299$\pm$0.005\\
$R_C$		&12.925$\pm$0.005\\
$I_C$		&12.511$\pm$0.004\\
$B-V$		&0.638$\pm$0.005\\
$V-R_C$		&0.374$\pm$0.007\\
$V-I_C$		&0.788$\pm$0.006\\
$V-K_S$		&1.770$\pm$0.022\\
$R_C-I_C$	&0.414$\pm$0.006\\
$I_C-K_S$	&0.982$\pm$0.021\\
$J-H$		&0.329$\pm$0.028\\
$J-K_S$		&0.382$\pm$0.032\\
$H-K_S$		&0.053$\pm$0.026\\
\hline                                   
\end{tabular}
\end{table}

Our $B$ and $V$ calibrated magnitudes are slightly fainter than the ones 
from \cite{everett2012}. We attribute the small differences ($\Delta B$=0.073, $\Delta V$=0.019),
to stellar activity variations, as seen in Fig.~\ref{fig:detrendKepler_light_curve}. In the Kepler
light curve these variations are of $\Delta m_{Kp}\sim$0.022, which is very close to the variation
we see in $V$ band. The $B-V$ index obtained by \cite{everett2012} is 0.584$\pm$0.030, which is slightly
bluer than ours, mainly as a consequence of the difference in the $B$ measurements.

\subsection{Effective temperature}
\label{subsec:Teff}

We derived the effective temperature of the system, $T_{\rm eff}$, using the empirical and model dependent,
temperature--color relations listed in Table~\ref{tab:Teff_calibrations}, our $BVR_C I_C$ colors derived
in Sect.~\ref{subsec:photcal_colors}, and the colors of the system published in 2MASS \citep{skrutskie2006},
and the SDSS \citep{eisenstein2011}. All those colors were measured outside of eclipses,
which guarantees that no systematics were introduced by the dimming of the system during eclipse events.
We still have the effect of stellar variability, which results in small variations of the color indexes.
Those variations are accounted for in the $T_{\rm eff}$ errors reported in Table~\ref{tab:Teff_calibrations}.

Table~\ref{tab:Teff_calibrations} summarizes the values of the $T_{\rm eff}$ derived using each
temperature--color relation. We assumed negligible reddening in the direction of T-Cyg1-12664 when
computing the colors but, in any case, the effect of the reddening over the distances involved will
be smaller than the uncertainties in colors among the optical and IR bands, as a consequence of
those colors being measured in different epochs and the intrinsic photometric variability of the system.
The mean effective temperature adopted for the system, computed as the average of all the results in
Table~\ref{tab:Teff_calibrations}, is 5560$\pm$160 K. This value is within 1$\sigma$ of the value
of $T_{\rm eff}$ obtained from fitting our spectra in Sect.~\ref{sec:RVobservations}, and corresponds
to a spectral type $G6V$, based on the tabulated values of \cite{mamajek2015}. We adopted this
result as the spectral type for the primary.

\begin{table}
\caption{\label{tab:Teff_calibrations}Mean effective temperature estimations resulting from our
photometric calibration, 2MASS and SDSS photometry and our spectroscopy for T-Cyg1-12664.
The interestellar reddening was not taken into account for this computation.}
\centering
\scriptsize
\begin{tabular}{lll}
\hline
\hline
Calibrations		&Color indices			&$T_{\rm eff}$ (K)\\
\hline
\multicolumn{3}{c}{Empirical}\\

\cite{cox2000}		&$(B-V)$, $(R-I_C)$		&5490$\pm$350\\
\cite{cox2000}\tablefootmark{a} &$(V-K_S)$, $(J-H)$	&5610$\pm$200\\
\cite{arribas1988}	&$(V-K_S)$			&5260$\pm$60\\
\cite{ballesteros2012}	&$(B-V)$			&5820$\pm$19\\
\cite{masana2006}	&$(V-K_S)$			&5460$\pm$15\tablefootmark{b}\\

\cite{ivezic2008}	&$(g'-r')$			&5527$\pm$44\\
\cite{fukugita2011}	&$(g'-r')$			&5564$\pm$28\\
\multicolumn{3}{c}{Model dependent}\\

\cite{bessell1979}	&$(B-V)$, $(V-I_C)$, $(R_C-I_C)$ &5420$\pm$300\\
\cite{houdashelt2000}	&$(B-V)$, $(V-R_C)$, $(V-I_C)$, &5680$\pm$230\\
			&$(V-K_S)$, $(J-K_S)$, $(H-K_S)$ &\\
\cite{lejeune1998}	&$(B-V)$, $(V-I_C)$,		&5570$\pm$290\\
			&$(V-K_S)$, $(R_C-I_C)$,	&\\
			&$(J-H)$, $(J-K_S)$, $(H-K_S)$	&\\
\hline
This work		&Spectroscopy			&5750$\pm$250\\
\hline
Mean value (adopted)			&&5560$\pm$160\\
\hline
\end{tabular}
\tablefoot{\\
\tablefoottext{a}{Interpolation from Table~7.6.}
\tablefoottext{b}{For $\log g$=4.5, $[M/H]$=-0.5).}
}
\end{table}

Our $T_{\rm eff}$ and spectral type of the primary do not agree with those derived by
\cite{devor2008thesis} and \cite{cakirli2013}. In those studies the authors claim a K5V spectral
type for the system and a mean $T_{\rm eff}$=4320$\pm$100 K. The main difference between their
calculations and ours is their use of B and V magnitudes from the USNO catalog and GSC2.3.2, which we
noticed in Sect.~\ref{subsec:photcal_colors}, are unreliable.

\subsection{Spectral energy distribution}
\label{subsec:SED}

\begin{figure}
\centering
\includegraphics[width=\hsize]{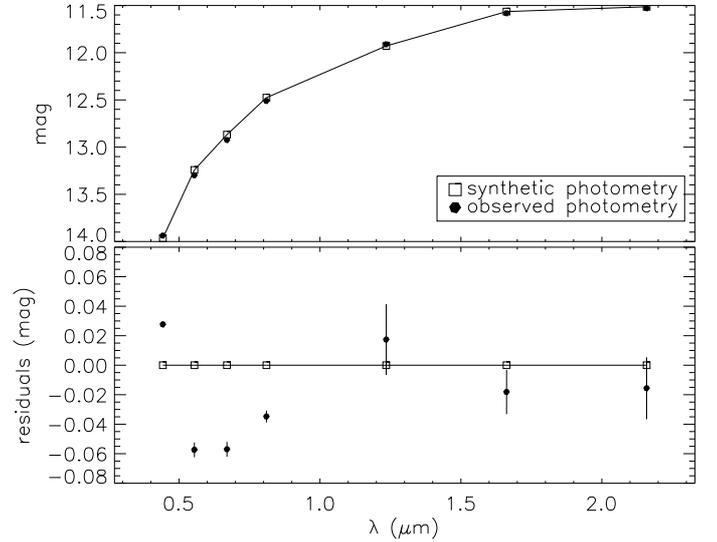}
  \caption{Comparison between the absolute optical and 2MASS photometry
  with the synthetic photometry obtained for a Kurucz model with $T_{\rm eff}$=5500 K,
  $\log g$=4.5, and $[M/H]$=-0.5. The uncertainties in the upper panel are smaller than
  the point sizes.}
      \label{fig:TCyg1_SED}
\end{figure}

Given the difference between the spectral type obtained in this work and those in the literature, and
as an additional check to the absolute photometry values derived in Sect.~\ref{subsec:photcal_colors},
we compared the optical and near-infrared colors measured for T-Cyg1-12664
to a \cite{kurucz1993} model template with T$_{eff}$=5500 K, $\log g=4.5$, and [M/H]=-0.5.
The result is shown in Fig.~\ref{fig:TCyg1_SED}, where the fit between the model and the observations
produces an rms of 0.037 mags. We also repeated the same comparison with the calibrated photometry in
Table~\ref{tab:calibrated_phot} and a template model with the same metallicity and $\log g$,
but with $T_{\rm eff}$=4250 K from \cite{cakirli2013}. The best fit in this case produces an
rms of 1.029 mags, with a clear systematic trend. Based on these results we conclude that the value
of 5560$\pm$160 K is a good estimate of the actual $T_{\rm eff}$ for T-Cyg1-12664.

\subsection{Rotation and synchronicity parameter}
\label{subsec:rotation}

The Kepler raw light curves of the system (see, for example, Fig.~\ref{fig:detrendKepler_light_curve}) reveals
periodic photometric variations owing to spots, and with a period clearly different from that of the binary.
In addition, and given the large differences in luminosity between the primary and secondary stars,
we can attribute those photometric variations to spots on the primary star. Therefore, we can measure
the rotation period of the primary, assuming that the spots rotate jointly with the photosphere.
After normalizing each quarter of the Kepler raw light-curve, and removing the eclipses using the ephemeris
equation of the system (see Sect.~\ref{subsec:ephemeris}), we calculated Lomb-Scargle periodograms \citep{scargle1982}
for each normalized light curve. A prominent peak is seen in the periodogram from which we
obtained rotation periods in a range between 3.840778 d and 4.250872 d.
We attribute those variations to changes in the different spot configurations over the time span of
the observations, and adopted a mean rotation period of $P_{rot}=3.968\pm0.091$ d. That rotation period
results in a synchronicity parameter $F=\omega_{rot}/\omega_{orb}=1.041\pm0.024$, using the period of
the binary computed in Sect.~\ref{subsec:ephemeris}. This value of $F$ is too small to produce a significant
deviation in the computed stellar radius and lets us assume that the primary star is rotationally
synchronized with the orbital period. This $P_{rot}$ must be taken as an estimate, given that the star
is not a rigid body with a well defined rotation period and, thus, this period depends on the latitude
of the spots. For each of the two main spots in the Q6 interval, we measured the times $t_{min}$ of each
associated minima and plotted the difference $T_0-t_{min}$ versus the sequential number of each minima,
with $T_0$ defined as the time of a primary eclipse. We fit a straight line to each spot's dataset and
obtained slopes of 3.9659$\pm$0.0092 d and 3.988$\pm$0.010 d, which confirms the presence of
differential rotation, as already revealed by the observed range of rotational periods.
In addition, the spot with the lower rotation period
is the one with lower latitude in the Q6-1 epoch (see Sect.~\ref{subsec:final_solutions}),
which suggests that the primary has solar-type differential rotation
\citep[see, for example,][ and references therein]{reinhold2015, lanza2014}.

\subsection{Third light}
\label{subsec:thirdlight}

T-Cyg1-12664 has an optical companion at a distance of 4.''6 as pointed by \cite{cakirli2013},
which is clearly seen in our images. We call this contaminant star T-Cyg1-C in the rest
of the text. The aperture of our ground-based photometry avoids contamination by light from this star.
Thus, our $VR_CI_C$ differential photometry is free of this third light contribution.
However, Kepler uses a photometric aperture of $\sim10-11$ pix for T-Cyg1-12664 (roughly 3x3 pix),
depending on the season, with a plate scale of 3''98/pix. Given that this contaminant is not listed in the
Kepler Input Catalog (KIC) it wouldn't have been accounted for during Kepler's third light
calibration, and the Kepler light curve of T-Cyg1-12664 is affected by the flux of T-Cyg1-C.

To estimate the amount of third light in the Kepler passband, we had to rely on the flux of the
contaminant in $V$, $R_C$, and $I_C$. We applied the photometric calibration performed on the night
of 2012-07-26 to T-Cyg1-C, to obtain its magnitudes and colors listed in Table~\ref{tab:T-Cyg1-C}.
From those colors we adopted $T_{\rm eff}=3761\pm56$ K for T-Cyg1-C, based on
the calibrations of \cite{cox2000}, \cite{bessell1991,bessell1979} and \cite{lejeune1998}. We didn't use
the $B-V$ color index since their $T_{\rm eff}$ is not consistent with the results of the other color indices
and the $B$-band magnitude is too faint. Thus, we rely only on the colors that make use of $V$, $R_C$, and $I_C$
photometry. From the obtained $T_{\rm eff}$, and assuming it is a main-sequence star, we adopted a spectral
type of M0V for this star, using the tables in \cite{mamajek2015}.

\begin{table}
\caption{\label{tab:T-Cyg1-C}Fluxes and colors for T-Cyg1-C, the contaminant in the Kepler passband.}
\centering
\begin{tabular}{lc}
\hline\hline
Parameter	&Value\\
\hline
B	&18.600+/-0.039\\
V	&17.822+/-0.029\\
R	&16.975+/-0.020\\
I	&16.009+/-0.017\\
\hline
$B-V$	&0.778+-0.049\\
$V-R_C$	&0.847+-0.035\\
$V_C-I_C$	&1.813+-0.034\\
$R_C-I_C$	&0.966+-0.026\\
\hline
\end{tabular}
\tablefoot{The color index B-V is not used in the computation of the $T_{\rm eff}$ of the
contaminating star (see text).}
\end{table}

To estimate the Kepler band flux, $K_p$, for T-Cyg1-C we fit an spectral energy distribution (SED) using the observed
colors and a Kurucz model spectrum with $T_{\rm eff}$=3750 K and solar metallicity and obtain a
geometric dilution parameter of $(R_*/D)^2 = (5.65\pm0.14)\times10^{-22}$ over the integrated $VR_CI_C$ bands
\citep[see][]{moro2000}. The uncertainty was computed performing Monte-Carlo fits over 100 iterations.
We used the same Kurucz spectrum used in the synthetic SED, scaled by the computed dilution factor and
integrated over the Kepler response function\footnote{Avaliable at \url{http://keplergo.arc.nasa.gov/CalibrationResponse.shtml}}
to estimate the contaminant flux due to T-Cyg1-C.
The uncertainty was also computed with a Monte-Carlo analysis, as in the dilution factor case,
and the obtained value is $F^{K_P}=(5.88\pm0.15)\times10^{-9}$  erg cm$^{-2}$ s$^{-1}$.
For T-Cyg1-12664 we followed the same procedure but using a dilution factor of
$(R/D)^2=(2.7291\pm0.0082)\times10^{-21}$, which results 
in a Kepler flux of $F_{K}=(2.5006\pm0.0071)\times10^{-7}$ erg cm$^{-2}$ s$^{-1}$.

Using these fluxes for the binary and the contaminant, the third light ratio in the Kepler band results in

\begin{center}
$l_3^{K_p}=0.02297\pm0.00058$.
\end{center}

As a check of the feasibility of this result, we compared this value with the estimated amount of third
light in the $V$, $R_C$, and $I_C$ bands using images taken the night of 2010-08-28 at an orbital
phase of 0.285. The values obtained for the differences in magnitude between T-Cyg1-12664 and
T-Cyg1-C were $\Delta V=4.619\pm0.028$, $\Delta R_C=4.186\pm0.018$, and $\Delta I_C=3.463\pm0.013$,
which translate into third light ratios at each passband

\begin{center}
$l_3^V=0.0140\pm0.0004$; $l_3^{R_C}=0.0207\pm0.0003$; $l_3^{I_C}=0.0396\pm0.0005$.
\end{center}

These results are of the same order of magnitude as those obtained for the $K_P$ band.
No other nearby stars are observed in the IAC80 images and there are no signs of
apsidal motion in the eclipse timings (primary or secondary) in the time span of the
observations (see Sect.~\ref{subsec:ephemeris}). Thus, we assumed that there is no other significant
source of third light in the system and we adopted the value for the Kepler band as the only source.
Nonetheless, during the modelling of this system in Sect.~\ref{subsec:final_solutions} we ran tests
to ensure that no other source of third light affects the $BVR_C I_C$ bands, for example,
if T-Cyg1-12664 were a multiple system with a third object that is not resolved in our images.
All these tests yielded negative results.

\subsection{Orbital period analysis and ephemeris}
\label{subsec:ephemeris}

We calculated the central times of eclipse in both the Kepler detrended light curves and the IAC80 light curves.
To find the central times of eclipse in the Kepler light curve, we fit either third or fourth order polynomials
to the points in each eclipse. The timings of the Kepler eclipses have relatively large uncertainties ($\sim$3 min)
because the events are sparsely sampled, given the 29.4 minutes sampling rate of the long cadence Kepler data. 
In the case of the IAC80 light curves, which are better sampled, we fitted sixth order polynomials for both the
primary and secondary eclipses. Times for the secondary eclipses were determined using the $I_C$-band light curve
since it shows the deepest eclipses and allows a better determination of the minima.
All the eclipse times are listed in Table~\ref{tab:eclipse_times}.

To compute a new ephemerides equation for the system, we also added the eclipse times published by \cite{devor2008}
and \cite{cakirli2013}. When necessary, we converted the times of minima from HJD to BJD
using the method of \cite{eastman2010}. We also conservatively assigned an uncertainty of 0.01 days to the
eclipses of \cite{devor2008}, given the dispersion of their photometry. All data combined provided a total of
228 primary eclipses over a time baseline of eight years. We fit all the measurements by a straight line,
eliminating two of the Kepler observations, which presented long deviations caused by bad sampling of the eclipses.
The result of that fit is the following updated ephemerides equation:

\onltab{
\begin{table*}
\caption{Measured times of minima for T-Cyg1-12664. Full version of this table is avaliable in
electronic form at the CDS.}
\label{tab:eclipse_times}
\centering
\begin{tabular}{cccc}
\hline
\hline
BJD             &Error      &Minimum Type   &Reference\\
day             &day        &--             &--\\
\hline
2453177.81205	&0.00100	&P	            &1\\
2453181.93795	&0.00100	&P	            &1\\
2453206.71285	&0.00100	&P	            &1\\
...&\\
\hline
\end{tabular}
\tablebib{(1)~\citet{devor2008}; (2) IAC80 data; (3) Kepler data;
(4) \citet{cakirli2013}
}
\end{table*}
}

\begin{equation}\label{eq:ephemeris}
 MinI(BJD) = 2455415.61831(5) + 4.1287955(4) n
\end{equation}

\begin{figure}
\centering
\includegraphics[width=\hsize]{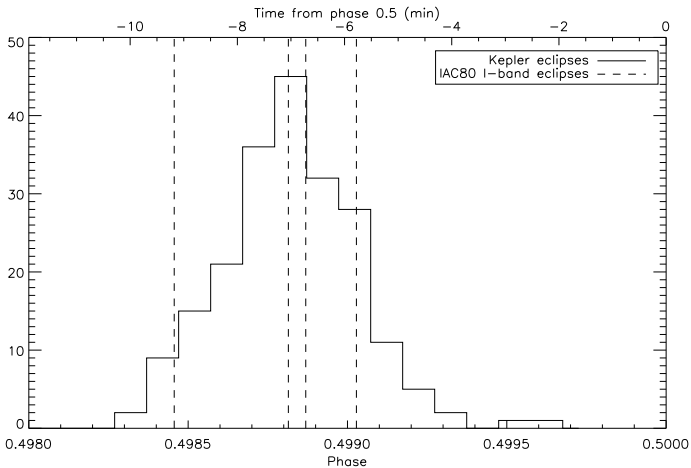}
  \caption{Histogram of the secondary eclipse distribution in phase (lower axis) and in time
  (upper axis). All the secondary eclipses are sistematically below phase 0.5. The vertical dashed
  lines show the position of the four IAC80 $I_C$-band secondary eclipses.}
      \label{fig:TCyg1_histo_secondaries}
\end{figure}

\noindent The residuals to this fit show no signs of deviation from a straight line, which suggest no
third body is present in the system, at least with orbital periods of the order of the obsevations timescale.
Finally, we show a histogram with the phase distributions of the secondary eclipses in Fig.~\ref{fig:TCyg1_histo_secondaries}.
All the eclipses are systematically below phase 0.5, with a mean phase of
$\Delta \phi=0.49894\pm0.00025$, which suggests that the binary is slightly eccentric.

\section{Radial velocity and light curve fits}
\label{RV_LC_fits}

\subsection{RV fitting procedure}
\label{subsec:RVfitting}

\begin{table}
\caption{RV fitting results.}
\label{tab:RVresults}      
\centering                          
\begin{tabular}{l r}        
\hline\hline                 
Parameter		&Value\\
\hline
\multicolumn{2}{c}{Adjusted Quantities}\\
\hline
$P$\tablefootmark{a} (d)	&4.1287955 $\pm$ $4\cdot10^{-7}$\\
$T_p$ (HJD)		&2454936.705 $\pm$ 0.011\\
$e$\tablefootmark{a}	&0.0365 $\pm$ 0.0014\\
$\omega$\tablefootmark{a} (deg)	&92.8 $\pm$ 2.2\\
$\gamma$ (km/s)		&-6.28 $\pm$ 0.53\\
$K_1$ (km/s)		&48.07 $\pm$ 0.74\\
$K_2$ (km/s)		&87.0 $\pm$ 2.5\\
\hline
\multicolumn{2}{c}{Derived Quantities}\\
\hline
$M_1\sin ^3i$ ($M_\odot$)	&0.677 $\pm$ 0.046\\
$M_2\sin ^3i$ ($M_\odot$)	&0.374 $\pm$ 0.017\\
$q = M_2/M_1$		&0.553 $\pm$ 0.018\\
$a_1\sin i$ ($10^6$ km)	&2.728 $\pm$ 0.042\\
$a_2\sin i$ ($10^6$ km)	&4.93 $\pm$ 0.14\\
$a  \sin i$ ($10^6$ km)	&7.66 $\pm$ 0.15\\
\hline
\multicolumn{2}{c}{Other Quantities}\\
\hline
$\chi^2$		&58.657\\
$\chi^2_{red}$		&1.128\\
$N_{obs}$ (primary)	&47\\
$N_{obs}$ (secondary)	&9\\
Time span (days)	&1507.86\\
$rms_1$ (km/s)		&5.079\\
$rms_2$ (km/s)		&3.492\\
\hline
\end{tabular}
\tablefoot{
\tablefoottext{a}{Parameter fixed beforehand.}
}
\end{table}

We fit our RVs jointly with the primary and secondary measurements from \cite{cakirli2013} and \cite{devor2008thesis}.
We fit a keplerian orbit to the data using the \verb;rvfit; code \citep{iglesias-marzoa2015}. This code is based on an
adaptive simulated annealing (ASA) algorithm and can simultaneously fit the complete set of keplerian parameters
$[P, T_p, e, \omega, \gamma, K_1, K_2]$ for the primary and secondary RV curves by $\chi^2$ function minimization.
The uncertainties in the parameters can be computed from the Fisher matrix or using
a Markov-Chain Monte Carlo method (MCMC). We chose this last method since it can deal with correlations
between parameters. MCMC was run over $10^6$ samples for each solution found with \verb|rvfit|.

We ran a first fit to obtain initial parameter values and to check the significance of an elliptical orbit.
In this first fit, we fixed only the orbital period to the value obtained in Sect.~\ref{subsec:ephemeris} and
left all the other parameters free. This first solution was compatible with a low eccentric, $e=0.069\pm0.016$,
orbit and a mass ratio of $q=0.558\pm0.018$. This elliptical orbit is confirmed by the small displacement
in phase of the secondary eclipses shown in Sect.~\ref{subsec:ephemeris}.

\begin{figure}
\centering
\includegraphics[width=\hsize]{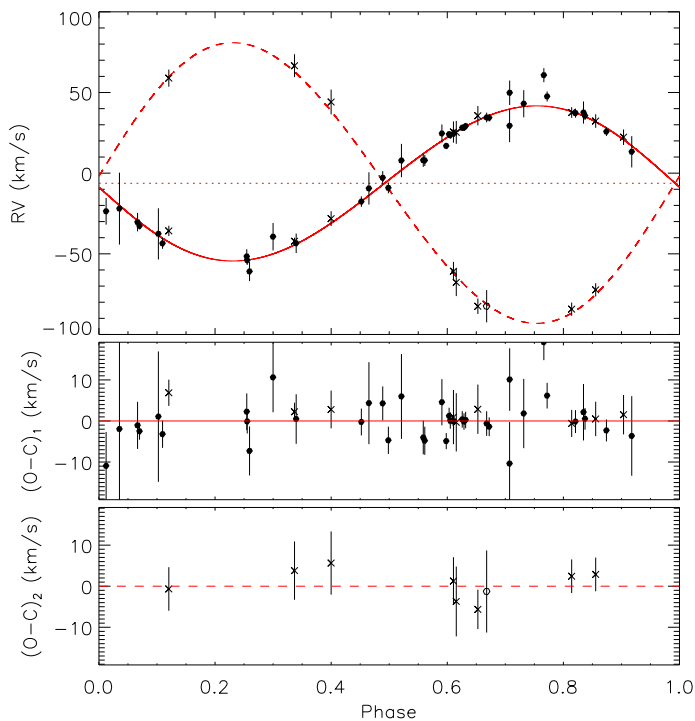}
  \caption{RV curve fit and residuals. The continuous and dashed curves are the fits for the
  primary and secondary stars, respectively. The filled and empty circles are the RV measurements
  for the primary and secondary star respectively. The crosses are the RV measurements from
  \cite{cakirli2013}. The dotted line in the upper panel is the value for $\gamma$. The phase
  was computed with Eq.~\ref{eq:ephemeris}.}
      \label{fig:RVcurve}
\end{figure}

The dispersion in the RV measurements and the sinusoidal appearance of the RV data
distribution can hide the subtle effect of a small eccentricity. To circumvent this limitation,
we decided to constrain the eccentricity and the argument of the
periastron using the Kepler light curves (see Sect.~\ref{subsec:JKTEBOP_analysis}) and fix
them to the values $e=0.0365\pm0.0014$ and $\omega=92.8\pm2.2$ deg. This new fit yielded the values in
Table~\ref{tab:RVresults}. The fitted RV curve is shown in Fig.~\ref{fig:RVcurve}.

\subsection{Light curves preliminary analysis with JKTEBOP}
\label{subsec:JKTEBOP_analysis}

We used the detrended Kepler data to obtain an initial model of the light-curve without spots and to fix some initial
parameters, namely $e\cos\omega$ and $e\sin\omega$. This allows us to start our combined analysis
of the multiband light curves with initial values near the final solution once the spots are included in the analysis.
For this goal we used JKTEBOP\footnote{JKTEBOP is written in FORTRAN 77 and the source code is available at
http://www.astro.keele.ac.uk/~jktcodes/jktebop.html} \citep{southworth2004}, which
is based on the EBOP code \citep{popper1981,etzel1981} and models the stars as triaxial ellipsoids.
We fixed $q$, $P$ and $T_0$ from the ephemeris Eq.~\ref{eq:ephemeris}, and the solution of the RV curves
(Sect.~\ref{subsec:RVfitting}). We fit the following parameters: the sum and ratio of fractional radii,
i.~e. $r_1+r_2$ and $r_1/r_2$, the orbital inclination $i$, the quantities $e\sin \omega$, $e\cos \omega$, the surface
bright of the secondary, and the light scale factor.

In this first stage, the effect of heating between the two components of the binary was not taken into account,
since it was eliminated  from the light curves when we detrended the out-of-eclipse variations. 
Therefore, the model coefficients associated with this effect were fixed to zero. In any case,
we ran tests to confirm this effect was not noticeable.
The third light contribution, $l_3$, was fixed to the value calculated in Sect.~\ref{subsec:thirdlight}.
Tests fitting $l_3$ yielded solutions with higher third light contributions, but poorer fits.
We adopted a square-root limb darkening (LD) law \citep{diaz-cordoves1992, vanhamme1993}, with the LD coefficients
interpolated from the tables provided by \cite{claret2011} and PHOENIX stellar models using the nominal $T_{\rm eff}$
of 5560$\pm$160 K and 3350$\pm$50 K for the primary and secondary components.
We adopted those values running preliminary PHOEBE tests with the optical IAC80 curves and the
RV solution for a circular orbit. The PHOENIX models were selected based in their lower $T_{\rm eff}$ limit,
although the LD coefficient were computed only for Z=0.0. The $\log g$ values were set
to 4.5 and 5.0 for each component, respectively. For the gravity-brightening coefficients we used values from
\cite{claret2011} for the same $T_{\rm eff}$.

To avoid problems caused by the long integration time of the Kepler data. we
split up each observation into six points to undergo a numerical integration. The effect of this Kepler
long integration time in light curves of eclipsing binaries \citep{coughlin2011} and exoplanet systems
\citep{kipping2010} is a known issue. These authors find that undersampling affects the morphological shape
of the eclipses (transits), in such a way that they appear to be shallower and have a longer duration.
This effect is most pronounced in eclipsing binaries with short periods and low sum of the fractional radii
($r_1+r_2$) values. T-Cyg1-12664, with a $r_1+r_2\simeq0.1$ is in the range where this effect can be noticed.

\begin{figure}
\centering
\includegraphics[width=\hsize]{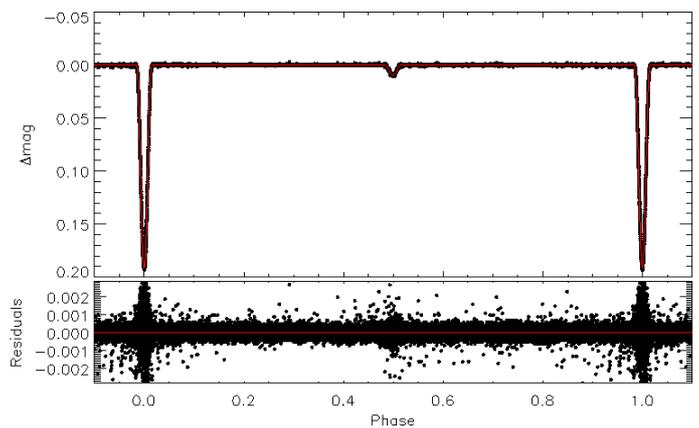}
  \caption{JKTEBOP fit to the Kepler detrended light-curve. A detailed plot of the
  fits in the eclipses is shown in Fig.~\ref{fig:keplerlc_jktebop_detail}}
      \label{fig:keplerlc_jktebop}
\end{figure}

\begin{figure}
\centering
\includegraphics[width=\hsize]{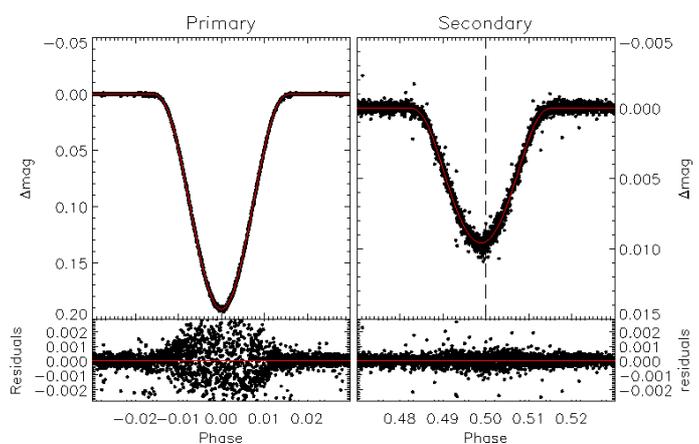}
  \caption{Detail of the Fig.~\ref{fig:keplerlc_jktebop} showing the fit in the eclipses. Note the
  change in the vertical scale in the two plots and the small displacement of the secondary eclipse from phase 0.5,
  indicated with a vertical dashed line.}
      \label{fig:keplerlc_jktebop_detail}
\end{figure}

The heavy detrending applied to the Kepler light curve produces points deviating largely from the model
in phases with strong light curve curvature near the center of the eclipses (see Fig.~\ref{fig:keplerlc_jktebop_detail}).
Thus, a conservative sigma-clipping of $10\sigma$ was applied in the fitting process to discard these points. With this
preliminary fit, we arrived at the solution shown in Figs.~\ref{fig:keplerlc_jktebop} and~\ref{fig:keplerlc_jktebop_detail},
where $i=86.87\pm0.14$, $e=0.0365\pm0.0014$, and $\omega=92.8\pm2.2$. The resulting fractional radii are 
$r_1=0.07282\pm0.00098$ and $r_2=0.03300\pm0.00050$. The uncertainties in the fitted parameter were computed using
a bootstrap analysis with 5 000 synthetic datasets.

\subsection{Final light curve analysis with PHOEBE}
\label{subsec:PHOEBE_analysis}
To properly model the multiband light-curves of T-Cyg1-12664, we used PHysics Of Eclipsing BinariEs
\citep[PHOEBE,][]{prsa2005, prsa2011}. 
Given the low value of $q$ for this system, the Roche lobe relative radii for each star
\citep{eggleton1983} are $r_{RL1}\simeq$0.43 and $r_{RL2}\simeq$0.33, well above the values obtained
for the relative radii from the preliminary fit with JKTEBOP, so we selected a detached eclipsing binary
model in PHOEBE.

\begin{figure}
\centering
\includegraphics[width=\hsize]{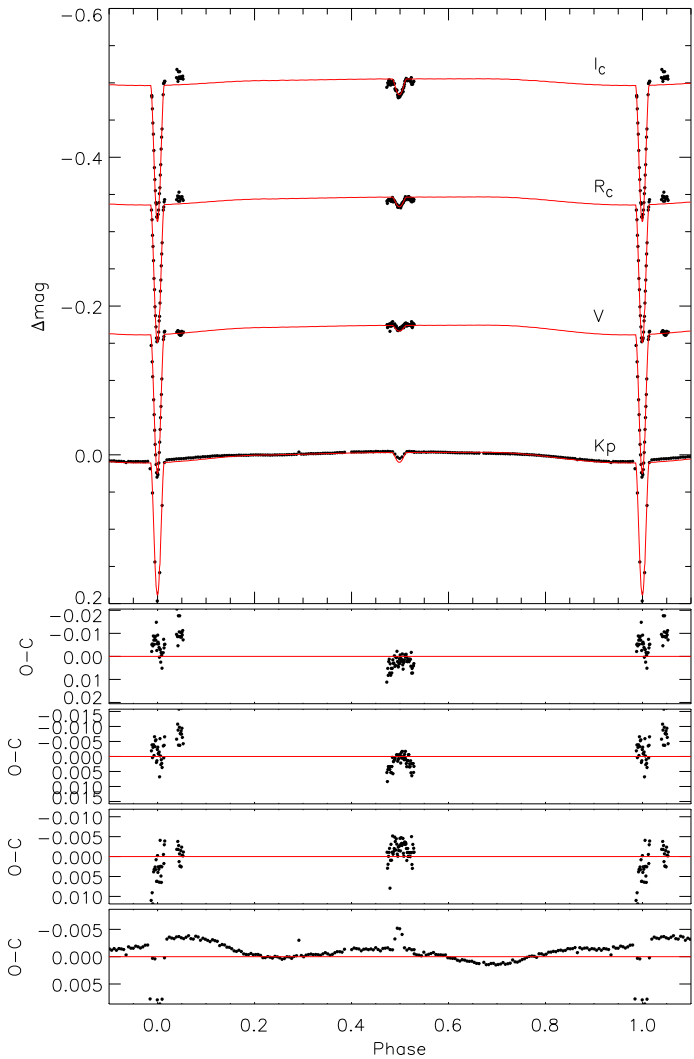}
  \caption{Light curves of the T-Cyg1-12664 in the epoch Q6-1 and fitted model with the
  parameters of the first column in Table~\ref{tab:LCresults}. From top to bottom, filters
  $I_C$, $R_C$, $V$, and Kepler band. The ground-based observations are displaced vertically
  to shrink the plot. The lower panels show the residuals of these fits in the same order.}
  \label{fig:cfase_modelos_Q6-1}
\end{figure}

\begin{figure}
\centering
\includegraphics[width=\hsize]{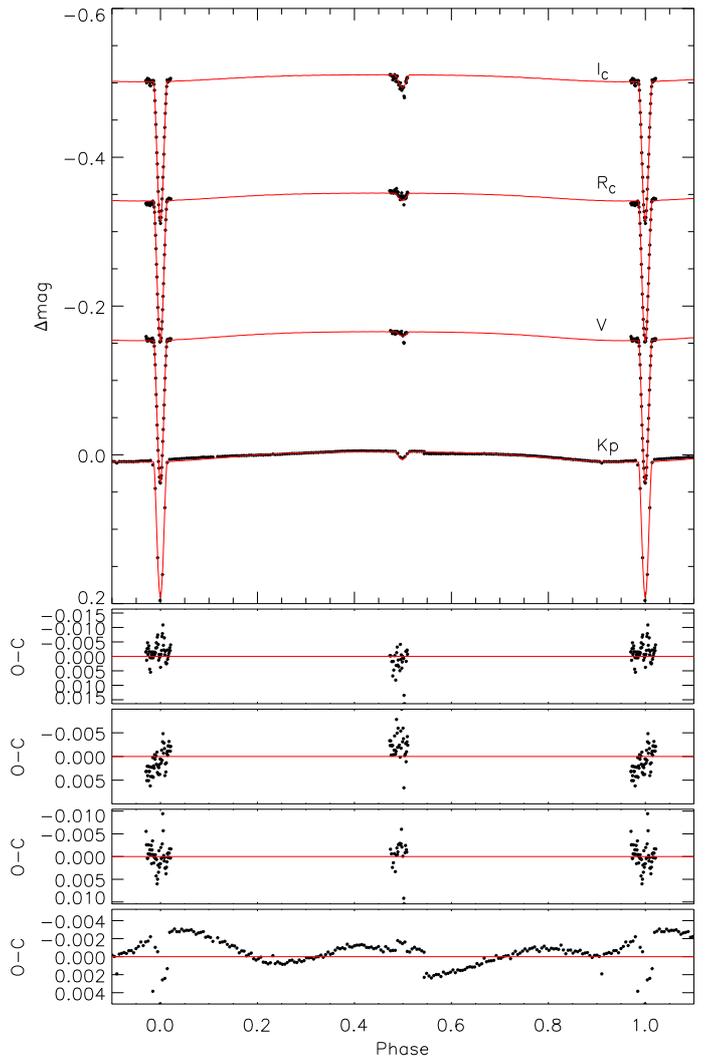}
  \caption{Same figure as \ref{fig:cfase_modelos_Q6-1}, but for the Q6-2 epoch. Note the
  jump in the Kepler photometry at phase 0.55 (see text).}
  \label{fig:cfase_modelos_Q6-2}
\end{figure}

We fit the IAC80 $VR_C I_C$ light curves simultaneously with the Kepler light curve, but only using
epochs of the Kepler light curve coeval with the ground-based observations.
This is a necessary step to ensure that we are modeling the same spot configuration.
Almost all IAC80 observations are coeval with Kepler's Q6. We identified three epochs in which a primary
and a contiguous secondary eclipses were observed simultaneously in all four bands. One epoch was discarded
owing to unidentified systematics in the IAC80 data, which affect the level of the primary eclipse. The other two
were selected for analysis with PHOEBE.

To avoid potential problems introduced by the $\simeq28$ min low-cadence duty cycle of the Kepler light-curve,
we performed a first fit using only the IAC80 light curves and the initial parameters derived from the detrended
Kepler light curve in Sect.~\ref{subsec:JKTEBOP_analysis} to get a first set of parameters for the binary
(radius, effective temperatures, etc). With these results, we ran a new fit, this time including the Kepler light
curve to properly model the spots. This two-step approach helps the fitting of the system because
1) the IAC80 $VR_C I_C$ light curves are best suited to constraining the effective temperatures, radii, and
inclination of the system because of their high cadence during eclipses and finer spectral resolution and
2) the Kepler lightcurve have a better coverage of the out-of-eclipse phases, which allows us to model the spots.
In the second fit, the physical and geometrical parameters of the binary are fixed and we only fit the spots.
We iterate this two-step process until a satisfactory model is obtained and variations in the resultant
parameters remain within the uncertainties for at least three consecutive fits. This approach has the
disadvantage of being slower than the simultaneous fit of all the  light curves but our preliminary test
showed that the fitting results were notably better from the residuals point of view.
The main advantage is that we retain both the physical information that are contained in the ground-based light
curves and the excellent time coverage of the spot configuration contained in the Kepler light curves.

{\renewcommand{\arraystretch}{1.2}
\begin{table}
\caption{T-Cyg1-12664 parameters computed for two different epochs.}
\label{tab:LCresults}
\begin{center}
{\scriptsize
\begin{tabular}{lcc}
\hline
\hline
Parameter		&Q6-1 epoch		&Q6-2 epoch\\
\hline
Time span		&2455376.51579 -	&2455380.70480 -\\
			&2455380.64350		&2455384.83250\\
			
\multicolumn{3}{c}{{\bf Geometric and orbital parameters}}\\

$P$ (d) (fixed)		&\multicolumn{2}{c}{4.1287955(7)}\\
$T_0$ (BJD) (fixed)	&\multicolumn{2}{c}{2455415.61831(5)}\\
$\Delta\phi$		&-0.000800$\pm$0.000024	&-0.000700$\pm$0.000023\\
$q$ (fixed)		&\multicolumn{2}{c}{0.553$\pm$0.018}\\
$\gamma$ (km s$^{-1}$) (fixed)	&\multicolumn{2}{c}{-6.28$\pm$0.53}\\
$i$ (deg)		&87.003$\pm$0.012	&86.931$\pm$0.010\\
$e$ (fixed)		&\multicolumn{2}{c}{0.0365$\pm$0.0014}\\
$a$ ($R_\sun$) (fixed)	&\multicolumn{2}{c}{11.03$\pm$0.21}\\
$\omega$ (deg) (fixed)	&\multicolumn{2}{c}{92.8$\pm$2.2}\\
$\Omega_1$		&14.391$\pm$0.057	&14.257$\pm$0.038\\
$\Omega_2$		&19.461$\pm$0.072	&18.937$\pm$0.061\\
$F_1$ (fixed)		&\multicolumn{2}{c}{1.041$\pm$0.024}\\
$F_2$ (fixed)		&\multicolumn{2}{c}{1.000}\\

\multicolumn{3}{c}{{\bf Fractional volumetric radii}}\\
$r_{1vol}$		&0.07240$\pm$0.00060	&0.07311$\pm$0.00040\\
$r_{2vol}$		&0.03039$\pm$0.00024	&0.03129$\pm$0.00022\\

\multicolumn{3}{c}{{\bf Radiative parameters}}\\

$T_{\rm eff1}$ (K) (fixed)	&\multicolumn{2}{c}{5560$\pm$160}\\
$T_{\rm eff2}$ (K)		&3540$\pm$160		&3380$\pm$160\\
Albedo (fixed)			&\multicolumn{2}{c}{0.5}\\
$\beta_1$, $\beta_2$	&0.4353, 0.3710		&0.4353, 0.3837\\
$l_3^{K_p}$ (fixed)	&\multicolumn{2}{c}{0.02297$\pm$0.00058}\\

\multicolumn{3}{c}{{\bf Limb-darkening coeficients (square-root law)}}\\

$x_1$,$y_1$ (bol)	&0.3811, 0.3636		&0.3811, 0.3636\\
$x_2$,$y_2$ (bol)	&-0.0427, 0.8284	&-0.0427, 0.8284\\
$x_1$,$y_1$ ($K_P$ band) &0.3598, 0.3869	&0.3598, 0.3869\\
$x_2$,$y_2$ ($K_P$ band) &0.0928, 0.8096	&0.0628  0.8721\\
$x_1$,$y_1$ ($V$ band)	&0.4786, 0.3242		&0.4786, 0.3242\\
$x_2$,$y_2$ ($V$ band)	&0.1759, 0.7896		&0.1544, 0.8349\\
$x_1$,$y_1$ ($R_C$ band) &0.3177, 0.4299	&0.3177, 0.4299\\
$x_2$,$y_2$ ($R_C$ band) &0.1592, 0.7278	&0.1061, 0.8169\\
$x_1$,$y_1$ ($I_C$ band) &0.1795, 0.4965	&0.1795, 0.4965\\
$x_2$,$y_2$ ($I_C$ band) &-0.1726, 1.0385	&-0.1848, 1.0833\\

\multicolumn{3}{c}{{\bf Spot 1 parameters (primary star)}}\\

Colatitude (deg)	&81.3$\pm$7.4		&19.51$\pm$0.74\\
Longitude (deg)		&18.2$\pm$1.5		&20.9$\pm$1.8\\
Radius (deg)		&19.97$\pm$0.40		&29.39$\pm$0.45\\
$T_{spot}/T_{surf}$	&0.98141$\pm$0.00087	&0.9686$\pm$0.0014\\

\multicolumn{3}{c}{{\bf Spot 2 parameters (primary star)}}\\

Colatitude (deg)	&34.9$\pm$4.1		&93$\pm$72\\
Longitude (deg)		&286.31$\pm$7.3		&291.8$\pm$9.0\\
Radius (deg)		&14.5$\pm$0.66		&8.1$\pm$1.5\\
$T_{spot}/T_{surf}$	&0.9801$\pm$0.0047	&0.9862$\pm$0.0037\\

\multicolumn{3}{c}{{\bf Parameters computed from MCMC}}\\

$\Delta\phi$		&-0.00027$^{+0.00028}_{-0.00024}$	&-0.00026$^{+0.00030}_{-0.00027}$\\
$i$ (deg)		&87.009$^{+0.025}_{-0.030}$		&86.9291$^{+0.0091}_{-0.0068}$\\
$\Omega_1$		&14.48$^{+0.48}_{-0.60}$		&14.30$^{+0.70}_{-0.42}$\\
$\Omega_2$		&18.97$^{+0.28}_{-0.38}$		&18.64$^{+0.45}_{-0.32}$\\
$T_{\rm eff2}$ (K)	&3548$^{+35}_{-38}$			&3365$^{+45}_{-52}$\\

\multicolumn{3}{c}{{\bf Fractional volumetric radii from MCMC}}\\

$r_{1vol}$		&0.0719$^{+0.0024}_{-0.0032}$		&0.0729$^{+0.0021}_{-0.0037}$\\
$r_{2vol}$		&0.0312$^{+0.0014}_{-0.0010}$		&0.0318$^{+0.0011}_{-0.0016}$\\

\multicolumn{3}{c}{{\bf Derived physical radii from MCMC}}\\

$R_{\rm 1}$ ($R_\sun$)	&0.793$^{+0.030}_{-0.038}$		&0.804$^{+0.028}_{-0.044}$\\
$R_{\rm 2}$ ($R_\sun$)	&0.344$^{+0.017}_{-0.013}$		&0.351$^{+0.014}_{-0.019}$\\

\hline
\end{tabular}
}
\end{center}
\end{table}
}

\begin{figure}
\centering
\includegraphics[width=\hsize]{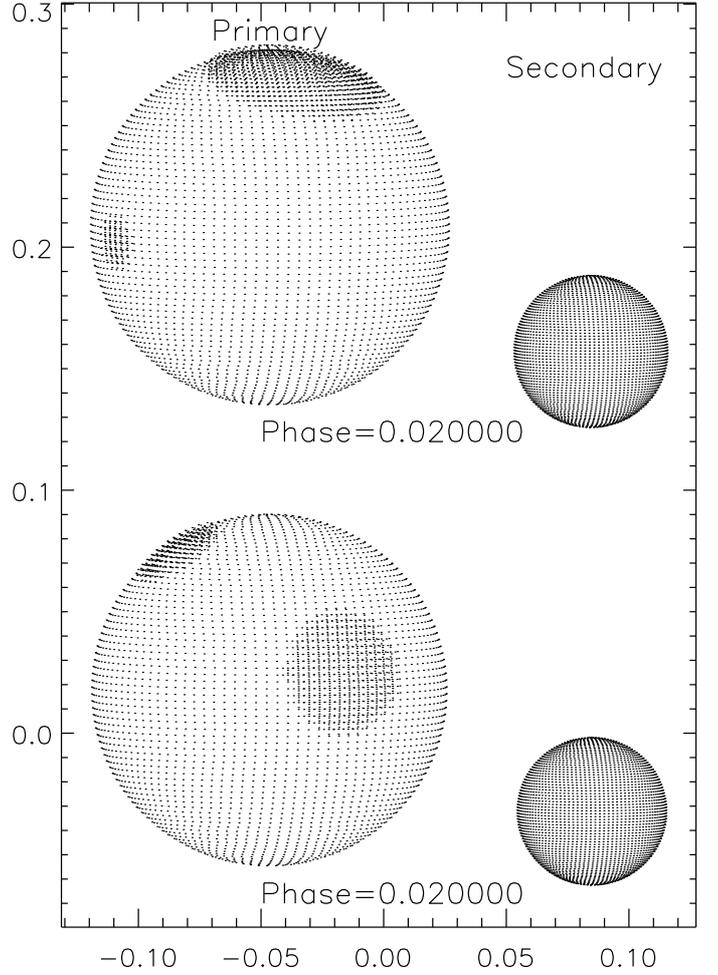}
  \caption{Representation of the spot configurations of T-Cyg1-12664 in the $(v,w)$ plane for
  the two epochs analyzed: the lower plot is for the Q6-1 epoch and the upper plot is for the
  Q6-2 epoch, which is displaced by an offset of 0.2 in the $w$ axis. Both axes have units of relative radius.
  The stars are represented at orbital phase 0.020, just after the primary eclipse, and
  the secondary star is orbiting towards the right side.}
      \label{fig:mesh_model}
\end{figure}

Given the large number of parameters in a complete binary system model, we tried to fix as many of
them as possible beforehand, using all available information. We fixed the period and $T_0$ to the
values in the ephemeris equation in Sect.~\ref{subsec:ephemeris}. The mass ratio $q$ and $\gamma$ were
fixed from the RV solution in Table~\ref{tab:RVresults}. The values for $e$ and $\omega$ were fixed
from the preliminary JKTEBOP analysis of the Kepler detrended light-curve in Sect.~\ref{subsec:Kepler_LC}.
The synchronicity parameter for the primary $F_1$ was set to 1.041$\pm$0.024, as obtained in Sect.~\ref{subsec:rotation}.
Although this is a very small value and it would not have practical effects on the shape of the star, we
included it for completeness. For the secondary star, the synchronicity parameter $F_2$ was assumed to be 1.0.

$T_{\rm eff1}$ was set to 5560$\pm$160 K, as computed in Sect.~\ref{subsec:Teff}, in agreement with
the spectroscopic value and the colors derived from the absolute photometry.
The albedos of the two components were set to $A_1=A_2=0.5$ as given by \cite{rucinski1969b},
since the temperatures and spectral types are compatible with those of stars with
convective envelopes. The values of the gravity-brightening exponents, $\beta_1$ and $\beta_2$, and those of the
LD coefficients were interpolated from \cite{claret2011} for each passband.
As in the case of the JKTEBOP preliminary light-curve analysis we used a square-root
LD law \citep{diaz-cordoves1992}. For the primary star these coefficient values were kept
fixed since $T_{\rm eff1}$ does not change throughout the fitting process. For the
secondary, we updated the coefficients values whenever necessary.
As in Sect.~\ref{subsec:JKTEBOP_analysis}, we used the same set of PHOENIX-based LD
coefficients. This is because, during the fitting process, the $T_{\rm eff2}$ evolves below the 3500 K limit
of the Kurucz LD tables. This occurred for all fitting attempts using the
Phoebe(2010) or \cite{vanhamme1993} LD coefficients, pointing to a $T_{\rm eff2}$ lower than 3500 K.
This situation is most evident in the MCMC computations (see Sect.~\ref{subsec:final_solutions}),
some of them computing the merit function with $T_{\rm eff}$ well below the 3500 K limit for the secondary.

The third light in the Kepler band was set
to the value computed in Sect.~\ref{subsec:thirdlight}.
For the $VR_C I_C$ band we set it to 0 since there is no evidence of a third body in the system
and our photometry excludes the nearby red star. Even so, we run preliminary tests to check this
assumption by fitting third light. The results were compatible
with $l_3$=0. For completeness, we also included mutual heating between the two components,
but its effect is negligible in this system.

\subsection{Spot modelling}

A good spot model is critical for this system.
Preliminary tests to decide on which component to place the spots showed that placing spots on the
secondary had negligible effects on the light curve. This was expected, given the large difference
in luminosity between the two components.

Our initial model consists of two spots on the primary located at phases $\phi\simeq-0.05$
and $\phi\simeq0.20$ ($\lambda\simeq18^\circ$ and $\lambda\simeq288^\circ$), and at a latitude of
$\sim45^\circ$, since this is the latitude that some studies claim is more affected by
spots \citep{hatzes1995,granzer2000}. Our final spot solutions for the two epochs
(see Figs.~\ref{fig:cfase_modelos_Q6-1} and~\ref{fig:cfase_modelos_Q6-2}) show small deviations in
the residuals with phase widths less than 0.5, so it cannot be produced by a single stellar spot.
We tried to reduce these deviations by adding more spots to the primary without seeing significant
improvements in the results. The Kepler light curve for the Q6-2 epoch shows a jump near phase
$\phi\simeq-0.55$, which prevents a better spot model fit, this jump is caused by small
uncorrected drifts in the Kepler photometry, by the evolution in size or temperature of spots or by a
combination of the two effects.

\subsection{Final solutions}
\label{subsec:final_solutions}

Taking into account the constraints and parameter values derived in the previous sections, the parameters
fitted in the final model were the phase shift $\Delta \phi$, the secondary temperature $T_{\rm eff2}$,
the orbital inclination $i$, the two surface potentials $\Omega_1$, $\Omega_2$, the passband luminosities,
and the spot parameters. The parameters common to the two columns were kept fixed during the fits and are the
same for both epochs. For $P$ and $T_0$, the uncertainty is shown within parentheses.
The final solutions for the two epochs analyzed are summarized in Table~\ref{tab:LCresults}, while the fitted
light curves and their residuals are shown in Figs.~\ref{fig:cfase_modelos_Q6-1} and \ref{fig:cfase_modelos_Q6-2}.

\begin{figure} 
\centering
\includegraphics[width=\hsize]{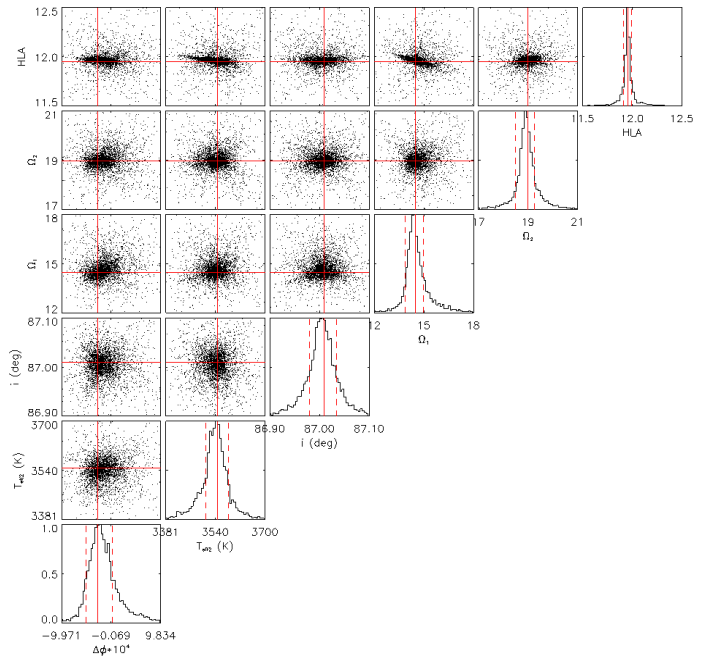}
  \caption{Parameter correlations resulting from MCMC fit and histograms of individual parameter distributions for the Q6-1 epoch
  light curve. The red lines show the values adopted in this work from the maximum of the histograms. Dashed vertical lines
  indicated the 68.3\% confidence interval.}
\label{fig:cornerQ6-1}
\end{figure}

\begin{figure} 
\centering
\includegraphics[width=\hsize]{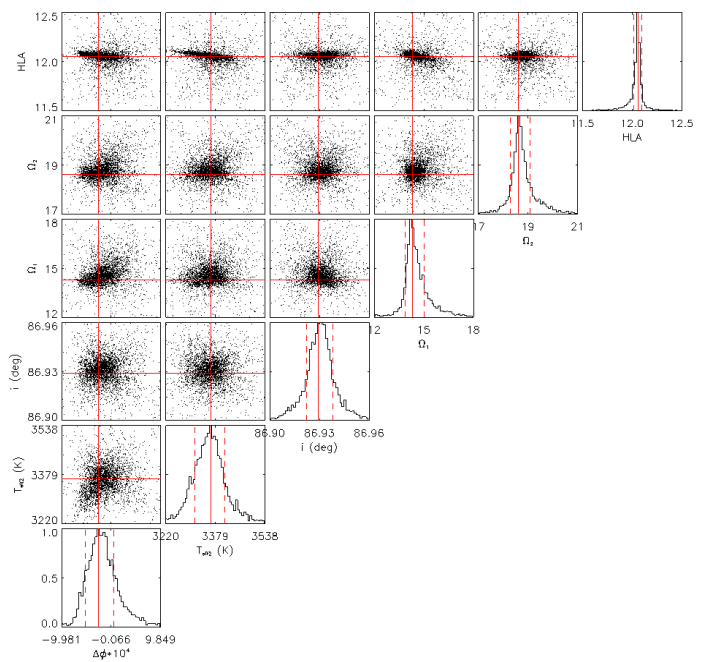}
  \caption{Same as Fig.~\ref{fig:cornerQ6-1} but for the Q6-2 epoch light curve.}
\label{fig:cornerQ6-2}
\end{figure}

Aditionally, we used an MCMC wrapper for PHOEBE to re-fit the parameter values
and to obtain a better estimation of the uncertainties in those parameters. The results of this computation
are shown at the end of Table~\ref{tab:LCresults}, while Figures~\ref{fig:cornerQ6-1} and \ref{fig:cornerQ6-2}
show the parameter correlations from MCMC simulations and histograms of individual parameter distributions.
To ensure the robustness of the fitted light curve model we ran a Gelman-Rubin diagnostics \citep{gelman1992}
on the fitted parameters. Table~\ref{tab:gelmanrubintest} shows the $\hat{R}$ values, all of them very near 1.0,
which indicate a robust solution.

\begin{table}
\caption{Computed $\hat{R}$ values from a Gelman-Rubin test for the fitted parameters in the two epochs analyzed.}  
\label{tab:gelmanrubintest}      
\centering                          
\begin{tabular}{l c c}        
\hline\hline                 
Fitted parameter	&Q6-1		&Q6-2\\    
\hline                                   
$\Delta\phi$	&1.01453	&1.02198\\
$T_{\rm eff2}$	&1.02718	&1.01771\\
$i$		&1.01481	&1.00577\\
$\Omega_1$	&1.01348	&1.01559\\
$\Omega_2$	&1.02770	&1.01063\\
HLA		&1.00799	&1.01104\\
\hline                                   
\end{tabular}
\end{table}

The fractional radii of each star are similar between the two epochs, but with differences slightly
larger than the computed uncertainties. The departure from a perfectly spherical shape is very
small ($\simeq$0.07\% for the primary and $\simeq$0.013\% for the secondary), as expected from the wide separation
between the stars and their small masses. The position of the spots differs significantly between epochs,
being the most stable parameter the longitude, which is tightly constrained by the light curve dips.
The differences between models for $T_{\rm eff}$ and the radii could be related to this variation in the spots
configuration, as noted by \cite{morales2010}. In Fig.~\ref{fig:mesh_model}, we
show a representation of the spot configuration resulting from our analysis in the two epochs analyzed.

\section{The system of T-Cyg1-12664}
\label{sec:system_TCyg1}

\subsection{Absolute parameters}
\label{subsec:absparams}

From the results in Tables~\ref{tab:RVresults} and \ref{tab:LCresults}, we computed the absolute parameters for the
components of T-Cyg1-12664. Those results are summarized in Table~\ref{tab:AbsDimensions}. The adopted radii, $T_{\rm eff2}$
and orbital inclination were computed from the mean for the two fits and their uncertainties are the standard deviations.
The secondary is an M3V star, based on its effective temperature \citep{mamajek2015}.
For the solar values we used the recommended IAU values $T_{\rm eff\sun}$=5772 K, $\log g_\sun$=4.438,
$M_{bol\sun}$=4.74 \citep{IAU_B2_2015,IAU_B3_2015}. The bolometric corrections were computed using the BC scale by
\cite{flower1996} \citep[but see][Sect.~2]{torres2010}.

\begin{table}
\caption{\label{tab:AbsDimensions}Absolute dimensions and main physical parameters of the
T-Cyg1-12664 system components.}
\centering
\begin{tabular}{lccc}
\hline\hline
Parameter			&Primary		&Secondary\\
\hline
Spectral type			&G6V			&M3V\\
Mass ($M_\sun$)			&0.680$\pm$0.045	&0.376$\pm$0.017\\
Radius ($R_\sun$)		&0.799$\pm$0.017\tablefootmark{a}	&0.3475$\pm$0.0081\tablefootmark{a}\\
$\log g$ (cgs)			&4.465$\pm$0.034	&4.931$\pm$0.028\\
$T_{\rm eff}$ (K)		&5560$\pm$160		&3460$\pm$210\tablefootmark{a}\\
$v_{rot}\sin i$ (km s$^{-1}$	&10.17$\pm$0.32\tablefootmark{b} &-\\
$v_{sync}\sin i$ (km s$^{-1}$)	&9.77$\pm$0.21\tablefootmark{c} &4.25$\pm$0.10\tablefootmark{c}\\
$BC_V$ (mag)			&-0.124$\pm$0.037	&-2.45$\pm$0.78\\
$L/L_\sun$			&0.550$\pm$0.068	&0.0156$\pm$0.0039\\
$M_{bol}$ (mag)			&5.40$\pm$0.13		&9.26$\pm$0.27\\
$M_V$ (mag)			&5.52$\pm$0.14		&11.71$\pm$0.83\\
\hline
\end{tabular}
\tablefoot{
\tablefoottext{a}{Taken as the average of the MCMC values measured at the two epochs.}
\tablefoottext{b}{Computed from the spots rotation period.}
\tablefoottext{c}{Projected rotational velocity expected for synchronous rotation and
a circular orbit.}
}
\end{table}

The uncertainties in the masses are 6.7\% for $M_1$ and 4.5\% for $M_2$, which are consistent with the
uncertainties in the RV amplitudes. The uncertainties in the radii are 2.1\% for $R_1$ and 2.3\% for $R_2$.
The radii are in good agreement with those obtained from the JKTEBOP model without spots
(see Sect.~\ref{subsec:JKTEBOP_analysis}), which resulted in $R_1$=0.803$\pm$0.019$R_\sun$ and
$R_2$=0.364$\pm$0.0089$R_\sun$. The difference in the secondary radius can be attributed to the strong impact of the
spots in the light curve, compared with the depth of the secondary eclipse.
The orbit is nearly edge-on with an inclination of $i$=86.969$\pm$0.056 degrees, adopted from the mean of the
two fits. The luminosity ratio in the visual band is $L_B/L_A=0.0034\pm0.0026$ and its uncertainty reflects
the uncertainty of the secondary absolute magnitude. Given the uncertainties in the $T_{\rm eff}$ scale,
we adopted uncertainties of $\pm$1 in the spectral types.

\cite{hut1981} showed that, in a eccentric orbit, the rotational angular velocity of
a star will tend to synchronize with the velocity at periastron as a consecuence of the strong
dependence of the tidal forces on distance, a condition called pseudosynchronization.
With the value of the orbital eccentricity, we computed the synchronicity parameter
for the case of pseudosynchronization \citep[][eq.~44]{hut1981} to be $F_{pseudo}=1.0765\pm0.0031$,
which is slighty larger than our measured syncronicity parameter $F=1.041\pm0.024$.
The difference could be due to differential rotation with spot latitude, in which case the spot's
rotation would not be representative of the actual stellar rotation. In addition, the
result from the Lomb-Scargle periodograms could be biased by the fact that there are, at least two
groups of spots, making the resulting rotation period a mean of the movement of the two groups.
Given the small eccentricity measured, we can consider that the orbit is circularized,
and this will impose a lower limit for the age of the system because of the long time required to
achieve this state. Using tidal friction theory \cite{zahn1977}, we find the time scale for synchronization
is $t_{sync}\simeq$6 Myr, and the time scale for circularization $t_{circ}\simeq$2 Gyr, which
sets a minimum age for the system.

\subsection{Distance, space velocities and age}
\label{subsec:age_distance_spacevel}

Combining the absolute V magnitudes of the two components, we obtain an absolute V magnitude for
T-Cyg1-12664 of $M_{Vtot}=5.52\pm0.14$. Given the apparent
visual magnitude of the system $V=13.299\pm0.005$ (see Sect.~\ref{subsec:photcal_colors}),
this gives an unreddened distance modulus of $m-M$=7.78, which translates into a distance
of $d=$360$\pm$22 pc.

We computed the $(U,V,W)$\footnote{Positive values of $U$, $V$, and $W$ indicate velocities
toward the Galactic center, Galactic rotation, and north Galactic pole, respectively.}
space velocities of the system using the \cite{johnson1987} algorithm, the distance computed above,
the system's systemic velocity (see Sect.~\ref{subsec:RVfitting}) and the 2MASS proper motion measurements
\citep{cutri2003}: $\mu_\alpha \cos{\delta}=-18\pm6$ mas yr$^{-1}$, $\mu_\delta=-6\pm2$ mas yr$^{-1}$.
The resultant velocities are $U=17.5\pm3.3$ km s$^{-1}$, $V=-11.1\pm1.7$ km s$^{-1}$, $W=11.5\pm5.9$ km s$^{-1}$,
and a total space velocity of $S=23.7\pm7.0$ km s$^{-1}$. These velocities place the binary near the border of the
area defined by \cite{eggen1989} as belonging to the young galactic disk. The velocities do not exactly agree with
the criteria of \cite{leggett1992} ($-50<U<20$, $-30<V<0$, $-25<W<10$, all in km s$^{-1}$)
but they lie very near the limit. Also the binary is not within any known early-type or late-type population tracer
\citep{skuljan1999}, although it lies near the V-middle branch for late-type stars ($U$=17.8, $V$=-15.37) km s$^{-1}$.
In addition, it cannot be related to any known moving group \citep{montes2001,maldonado2010}. Therefore, we cannot
impose constraints on the age of the system based on kinematic criteria. However, based on circularization and
synchronization times for a $P\simeq$4.13 days orbit, we can speculate on an age of at least 2 Gyrs for the system.
This age rules out the possibility of T-Cyg1-12664 being pre-main-sequence, and is a clue that its components
are main-sequence stars.

\begin{figure} 
\centering
\includegraphics[width=\hsize]{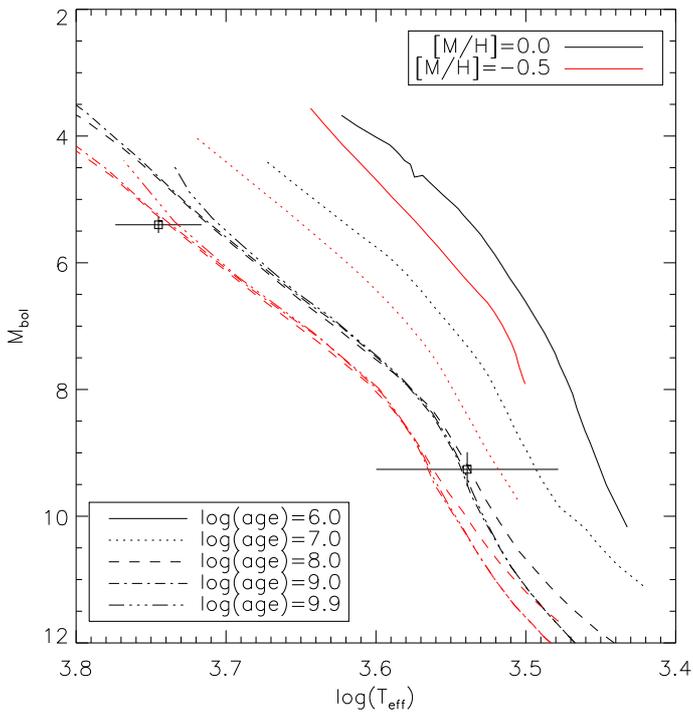}
  \caption{Position in the $M_{bol}$-$\log T_{\rm eff}$ diagram of the two components in the system of T-Cyg1-12664.
  BCAH98 isochrones are for [M/H]=0.0 (black) and [M/H]=-0.5 (red) for $\log(age)$=6.0, 7.0, 8.0, 9.0 and 9.9.}
\label{fig:Mbol_logTeff}
\end{figure}

In Fig.~\ref{fig:Mbol_logTeff}, we plotted the \cite{baraffe1998} (hereafter BCAH98) isocrones
for [M/H]=0.0 (in black) and for [M/H]=-0.5 (in red). All the isochrones older than log(age)=8.0 fit
equally well the components of this system. The primary is better fitted by isochrones for
[M/H]=-0.5, and the secondary is better fitted by isochrones for [M/H]=0.0. The effect
of the subsolar metallicity could explain the primary's effective temperature, but not its oversize.
However, the difference in the space parameter for the two sets of isochrones
for the two metallicities is smaller than the uncertainties in the computed stellar parameters,
and the uncertainty about the actual metallicity of the system remains.
A specific metallicity analysis for this system would resolve this question.
All that can be said is that this system is old and we can safely discard a young pre-main-sequence
system.

\subsection{Comparison with models and previous analyses}
\label{subsec:comparison_with_models}

\begin{figure}
\centering
\includegraphics[width=\hsize]{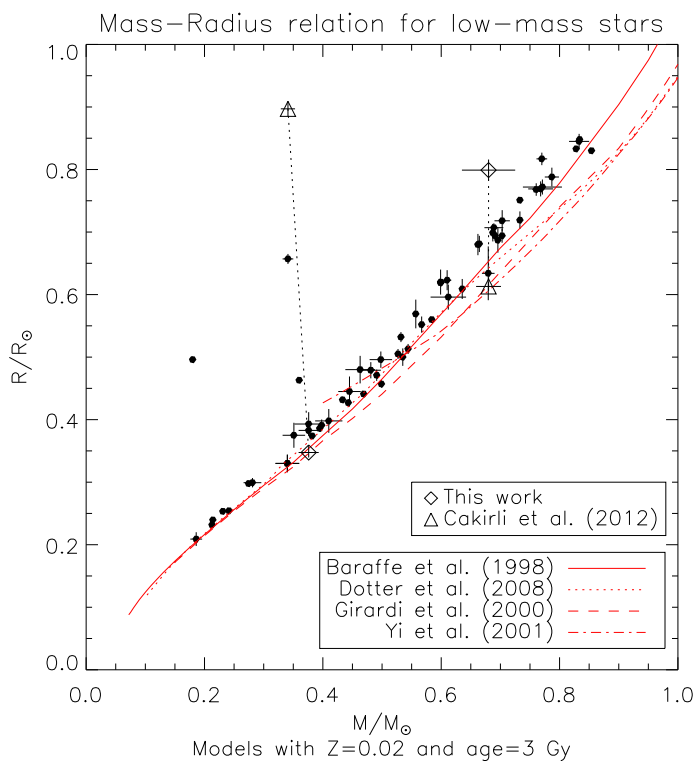}
  \caption{Mass-radius relations of stars between 0.2 and 1.0 $M_\sun$ predicted by four
  stellar models (see text). The filled circles represent well-known LMDEB, used as benchmarks
  for the models. The triangles represent the location of T-Cyg1-12664 in the \cite{cakirli2013}
  paper and the diamonds are our solution for this system. We connect the two solutions for
  the primary and the secondary with a dotted line to guide the eye to the difference in the
  physical parameters that arise from the two solutions. Metallicity and age are set for reference
  of the whole set of stars and they are not a fit to the T-Cyg1-12664 parameters.} \label{fig:M-R}
\end{figure}

\begin{figure}
\centering
\includegraphics[width=\hsize]{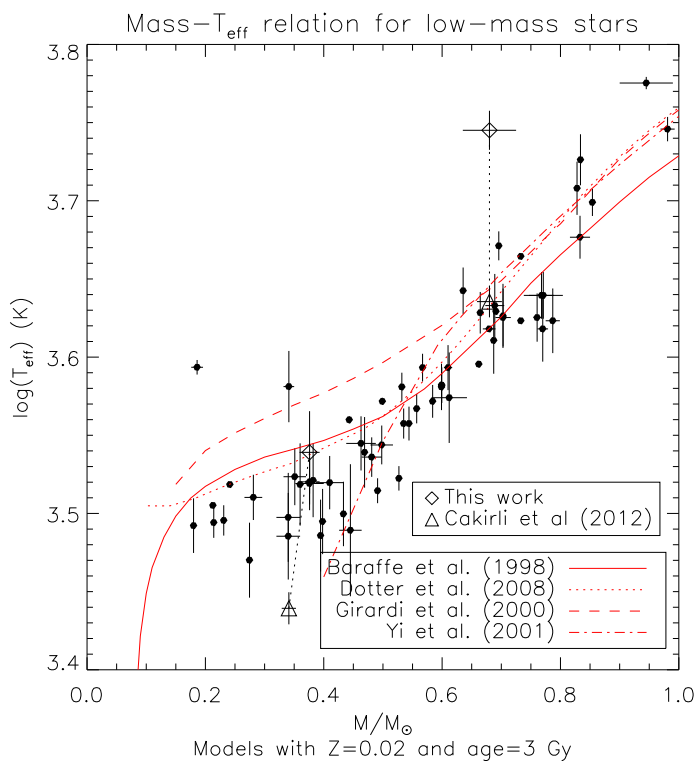}
  \caption{Mass-$\log T_{\rm eff}$ relations for the same LMDEB systems as in Fig.~\ref{fig:M-R}. Again,
  we connected with a dotted line the solutions derived by \cite{cakirli2013} for the primary and
  secondary with our solutions. Metallicity and age are set for reference
  of the whole set of stars and they are not a fit to the T-Cyg1-12664 parameters.} \label{fig:M-Teff}
\end{figure}

Low-mass stars in eclipsing binaries are commonly found to have larger radii and lower effective temperatures than
those predicted by current models. In Figs.~\ref{fig:M-R} and~\ref{fig:M-Teff}, we compare the mass, radius, and
temperature of the components of T-Cyg1-12664 to models and to other binaries. The models in the figures are those of
\cite{baraffe1998}, \cite{dotter2008}, \cite{girardi2000}, and \cite{yi2001}, all of them for solar metallicity
(Z=0.02) and an age of 3 Gyr, which cover the entire low-mas regime.
We adopt those values since they provide a good fit to the overall sample of mass-radius measurements of
main sequence low-mass stars in the literature, and also the secondary of T-Cyg-12664, although not the primary.
We note, however, that this model is used for illustration of the entire observational sample of low-mass stars,
and is not a fit to the T-Cyg-12664.

The mixing length parameter for the \cite{baraffe1998} model is $\alpha=l/H_p=1$.
The other binaries included in the figure (listed in Table~\ref{tab:eclipsing_binaries_literature}) are all the current
systems with masses and radii measured with relative precisions better than 7\%. We also included the
previous measurements by \cite{cakirli2013} for this system in the figures.

Our results show agreement between the parameters of the secondary and models, in particular with the predictions from
\cite{baraffe1998} and \cite{dotter2008}. This is a particularly important result, since the secondary has a mass very
close to the fully convective stellar boundary at $\sim$0.35 $M_\sun$. The parameters of the primary, on the other hand,
deviate significantly from the model predictions, both in the mass-radius and the mass-$\log{T_{\rm eff}}$ plots.
The primary is oversized $\sim$34\% over the value of the radius predicted by the models. This deviation is larger than
any other well measured binary components in this mass range. In the mass-$\log{T_{\rm eff}}$ plane, the primary exhibits
a very high effective temperature, being over 1000K hotter than what models for a star of that mass predict.
We must emphasize that the temperature has been measured directly from the photometric colors, which are found to be
consistent among photometric systems, and spectroscopy (see Sects.~\ref{subsec:Teff} and \ref{subsec:SED}), and is not
a result of the modeling process. At present, we can not find an explanation for this discrepancy,
but there is at least one other low-mass binary that exhibits this effect \citep[see][]{gomez2014}.

\onllongtab{
\begin{longtable}{lcccc}
\caption{List of benchmark low-mass components in eclipsing binaries taken from the literature. The
measured masses and radii have relative uncertainties less or equal 7\%.}
\label{tab:eclipsing_binaries_literature}\\
\hline 
\hline
Component of binary	&$M$ ($M_\sun$)		&$R$ ($R_\sun$)		&$T_{\rm eff}$ (K)	&References\\
\hline
YY Gem A/B		    &0.5992$\pm$0.0047	&0.6191$\pm$0.0057	&3820$\pm$100	&1\\
V818 Tau B		    &0.7605$\pm$0.0062	&0.768$\pm$0.010	&4220$\pm$150	&1\\
CU Cnc A		    &0.4333$\pm$0.0017	&0.4317	$\pm$0.0052	&3160$\pm$150	&2\\
CU Cnc B		    &0.3980$\pm$0.0014	&0.3908$\pm$0.0094	&3125$\pm$150	&2\\
GU Boo A            &0.610$\pm$0.007	&0.623$\pm$0.016	&3920$\pm$130	&3\\
GU Boo B	     	&0.599$\pm$0.006	&0.620$\pm$0.020	&3810$\pm$130	&3\\
CM Dra A		    &0.2310$\pm$0.0009	&0.2534$\pm$0.0019	&3130$\pm$70	&4\\
CM Dra B		    &0.2141$\pm$0.0010	&0.2396$\pm$0.0015	&3120$\pm$70	&4\\
IM Vir A		    &0.981$\pm$0.012	&1.061$\pm$0.016	&5570$\pm$100	&5\\
IM Vir B		    &0.6644$\pm$0.0048	&0.681$\pm$0.013	&4250$\pm$130	&5\\
2MASS J05162881+2607387 A &0.787$\pm$0.012   &0.788$\pm$0.015	&4200$\pm$200	&6\\
2MASS J05162881+2607387 B &0.770$\pm$0.009   &0.817$\pm$0.010	&4150$\pm$20	&6\\
T-Lyr1-17236 A	    &0.6795$\pm$0.0107	&0.634$\pm$0.043	&4150		    &7\\
T-Lyr1-17236 B	    &0.5226$\pm$0.0061	&0.525$\pm$0.052	&3700		    &7\\
LSPM J1112+7626 A   &0.3946$\pm$0.0023	&0.3860$\pm$0.0055	&3061$\pm$162	&8\\
LSPM J1112+7626 B   &0.2745$\pm$0.0012	&0.2978$\pm$0.0049	&2952$\pm$163	&8\\
GJ 3236 A	        &0.376$\pm$0.017	&0.3828$\pm$0.0072	&3313$\pm$110   &9\\
GJ 3236 B	        &0.281$\pm$0.015	&0.2992$\pm$0.0075	&3238$\pm$108   &9\\
V636 Cen B	        &0.854$\pm$0.003	&0.830$\pm$0.004	&5000$\pm$100   &10\\
NSVS 07394765 A		&0.360$\pm$0.005	&0.463$\pm$0.004	&3300$\pm$200   &11\\
NSVS 07394765 B		&0.180$\pm$0.004	&0.496$\pm$0.005	&3106$\pm$125   &11\\
NGC2204-S892 A		&0.733$\pm$0.005	&0.719$\pm$0.014	&4200           &12\\
NGC2204-S892 B		&0.662$\pm$0.005	&0.680$\pm$0.017	&3940$\pm$20    &12\\
RX J0239.1-1028 A	&0.733$\pm$0.002	&0.751$\pm$0.002	&4618$\pm$14    &13\\
RX J0239.1-1028 B 	&0.691$\pm$0.002	&0.694$\pm$0.003	&4258$\pm$14    &13\\
19b-2-01387 A		&0.498$\pm$0.019	&0.496$\pm$0.013	&3498$\pm$100   &14\\
19b-2-01387 B    	&0.481$\pm$0.017	&0.479$\pm$0.013	&3436$\pm$100   &14\\
19c-3-01405 A		&0.410$\pm$0.023	&0.398$\pm$0.019	&3309$\pm$130   &14\\
19c-3-01405 B    	&0.376$\pm$0.024	&0.393$\pm$0.019	&3305$\pm$130   &14\\
19e-3-08413 A		&0.463$\pm$0.025	&0.480$\pm$0.022	&3506$\pm$140   &14\\
19e-3-08413 B		&0.351$\pm$0.019	&0.375$\pm$0.020	&3338$\pm$140   &14\\
KOI-126 B	        &0.2413$\pm$0.0030	&0.2543$\pm$0.0014	&3300           &15\\
KOI-126 C	        &0.2127$\pm$0.0026	&0.2318$\pm$0.0013	&3200           &15\\
LP 133-373 A	        &0.340$\pm$0.020	&0.330$\pm$0.014	&3058$\pm$195   &16\\
LP 133-373 B	        &0.340$\pm$0.020	&0.330$\pm$0.014	&3144$\pm$206   &16\\
ASAS J011328–3821.1 Aa	&0.612$\pm$0.030	&0.596$\pm$0.020	&3750$\pm$250   &17\\
ASAS J011328–3821.1 Ab	&0.445$\pm$0.019	&0.445$\pm$0.024	&3085$\pm$300   &17\\
ASAS J045304-0700.4 A	&0.8338$\pm$0.0036	&0.848$\pm$0.005	&5324$\pm$200   &18\\
ASAS J045304-0700.4 B	&0.8280$\pm$0.0040	&0.833$\pm$0.005	&5105$\pm$200   &18\\
ASAS J082552-1622.8 A	&0.7029$\pm$0.0045	&0.694$\pm$0.011	&4230$\pm$200   &18\\
ASAS J082552-1622.8 B	&0.6872$\pm$0.0049	&0.699$\pm$0.014	&4080$\pm$200   &18\\
ASAS J093814-0104.4 A	&0.771$\pm$0.033	&0.772$\pm$0.012	&4360$\pm$150   &19\\
ASAS J093814-0104.4 B	&0.768$\pm$0.021	&0.769$\pm$0.012	&4360$\pm$150   &19\\
ASAS J212954-5620.1 A	&0.833$\pm$0.017	&0.845$\pm$0.012	&4750$\pm$150   &19\\
ASAS J212954-5620.1 B	&0.703$\pm$0.013	&0.718$\pm$0.017	&4220$\pm$180   &19\\
MG1-78457 A	        &0.527$\pm$0.002	&0.505$\pm$0.008	&3330$\pm$60    &20\\
MG1-78457 B	        &0.491$\pm$0.001	&0.471$\pm$0.009	&3270$\pm$60    &20\\
MG1-116309 A		&0.567$\pm$0.002	&0.552$\pm$0.013	&3920$\pm$80    &20\\
MG1-116309 B		&0.532$\pm$0.002	&0.532$\pm$0.008	&3810$\pm$80    &20\\
MG1-506664 A		&0.584$\pm$0.002	&0.560$\pm$0.004	&3730$\pm$90    &20\\
MG1-506664 B		&0.544$\pm$0.002	&0.513$\pm$0.008	&3610$\pm$90    &20\\
MG1-646680 A		&0.499$\pm$0.002	&0.457$\pm$0.006	&3730$\pm$20    &20\\
MG1-646680 B		&0.443$\pm$0.002	&0.427$\pm$0.006	&3630$\pm$20    &20\\
MG1-1819499 A		&0.557$\pm$0.001	&0.569$\pm$0.023	&3690$\pm$80    &20\\
MG1-1819499 B		&0.535$\pm$0.001	&0.500$\pm$0.014	&3610$\pm$80    &20\\
MG1-2056316 A		&0.469$\pm$0.002	&0.441$\pm$0.002	&3460$\pm$180   &20\\
MG1-2056316 B		&0.382$\pm$0.001	&0.374$\pm$0.002	&3320$\pm$180   &20\\
AK For A		        &0.6958$\pm$0.0010	&0.687$\pm$0.020	&4690$\pm$100   &21\\
AK For B		        &0.6355$\pm$0.0007	&0.609$\pm$0.016	&4390$\pm$150   &21\\
1SWASP J011351.29+314909.7 A &0.945$\pm$0.045	&1.378$\pm$0.058	&5961$\pm$54    &22\\
1SWASP J011351.29+314909.7 B &0.186$\pm$0.010	&0.209$\pm$0.011	&3922$\pm$42    &22\\

T-Cyg1-12664 A		&0.680$\pm$0.021	&0.613$\pm$0.007	&4320$\pm$100	&23\\
T-Cyg1-12664 B		&0.341$\pm$0.012	&0.897$\pm$0.012	&2750$\pm$65	&23\\

T-Cyg1-12664 A		&0.680$\pm$0.045	&0.799$\pm$0.017	&5560$\pm$160	&24\\
T-Cyg1-12664 B		&0.376$\pm$0.017	&0.3475$\pm$0.0081	&3460$\pm$210	&24\\
\hline
\end{longtable}
\tablebib{(1)~\citet{torres2002}; (2) \citet{ribas2003};
(3) \citet{lopez-morales2005}; (4) \citet{morales2009};
(5) \citet{morales2009b}; (6) \citet{bayless2006};
(7) \citet{devor2008b}; (8) \citet{irwin2011};
(9) \citet{irwin2009}; (10) \citet{clausen2009};
(11) \citet{cakirli2013b}; (12) \citet{rozyczka2009};
(13) \citet{brogna2006}; (14) \citet{birkby2012};
(15) \citet{carter2011}; (16) \citet{vaccaro2007};
(17) \citet{helminiak2012}; (18) \citet{helminiak2011};
(19) \citet{helminiak2011b}; (20) \citet{kraus2011};
(21) \citet{helminiak2014}; (22) \citet{gomez2014};
(23) \citet{cakirli2013}; (24) this work.}
}

Our results also differ from those of \cite{cakirli2013}, the main differences in the analysis being
the setting of a reliable set of color indexes, the treatment of the eccentricity, and the
processing of the Kepler light curves. These differences can explain the disagreement in the parameters
obtain from the light curves, i.e. the radii, the $T_{\rm eff}$, and luminosity-related quantities.
Only the masses, derived from the RV curves, are similar, although our secondary mass is slightly larger.
We attribute this to improvements in the RV curve of the primary and the detection of a slight
eccentricity for the system.

T-Cyg1-12664 was also analyzed by \cite{prsa2011b} using Eclipsing Binaries via Artificial Intelligence
\citep[EBAI,][]{prsa2008} on a survey of eclipsing binaries found in the Kepler field
of view. EBAI relies on trained neural networks to yield principal parameters for all the binary stars
in the catalog. In the accompanying table, we found the following values:
$T_{\rm eff}=$5669 K, $\log g$=4.308, $\sin i$=0.99441, $e\sin \omega$=0.04452, $e\cos \omega$=-0.35923,
$T_2/T_1$=0.5965, and $\rho_1+\rho_2$=0.10881, with an effective temperature in reasonable agreement
with the value obtained from our photometry.
The same authors published the Kepler Eclipsing Binary Catalog\footnote{Avaliable at \url{http://keplerebs.villanova.edu/v2}.}
(KEB) online. A query of this catalog using the Kepler name of the system (KIC 10935310) returns a different set of
physical parameters: $\sin i$=0.99479, $e\sin \omega$=0.05606, $e\cos \omega$=0.22161, $T_2/T_1$=0.78209,
and $\rho_1+\rho_2$=0.13246. The values for $e$ and $\omega$ arising from both solutions are incompatible with our
radial velocity and photometric curves.
However, a new analysis of the Kepler dataset by the Kepler EB team gives $\omega$=94.2 degrees and $e$=0.0372,
which are consistent with our values (Pr{\v s}a, priv. comm.).

\section{Conclusions}
\label{sec:conclusions}

Our analysis of the T-Cyg1-12664 low-mass eclipsing binary system shows a different picture from
other published analyses \citep{devor2008thesis, cakirli2013}. The main differences between
our analysis and the previous ones are: the measurement of consistent and accurate calibrated
magnitudes in the optical bands, which lead to a precise determination of the $T_{\rm eff}$ scale;
a careful treatment of the contamination by nearby stars in the Kepler band photometry,
with a direct impact in the orbital inclination and, thus, in the fractional radii; and the
discovery of a small displacement of the secondary eclipse which lead to the analysis
of the system as an eccentric binary.

Our resuls for this system show an oversized primary with $M_1=0.680\pm0.045$ $M_\sun$,
$R_1=0.799\pm0.017$ $R_\sun$, and $T_{\rm eff1}=5560\pm160$ K, in a slightly eccentric orbit
with $e=0.0365\pm0.0014$ and $i=86.969\pm0.056$. The secondary is a star near the fully convective stellar 
mass boundary with $M_2=0.376\pm0.017$ $M_\sun$, $R_2=0.3475\pm0.0081$ $R_\sun$, and $T_{\rm eff2}=3460\pm210$ K.
Its parameters agree well with models.

The faintness of the secondary component is a drawback in the measurement of radial velocities for
this system, which only shows the radial velocities of the primary with telescopes of medium size.
This prevents us from directly measuring $q$ for this system and forces us to rely on published
radial velocities for the secondary. The secondary star appears to have a mass in
a key region for models, i.e. the transition to fully convective stars, which remains poorly sampled
by observations. Therefore, new measurements of the secondary radial velocity curve of this system
using, for example, near-IR spectroscopy will be important.

\begin{acknowledgements}
We thank the referee A. Pr{\v s}a for helpful comments on the manuscript.
This article is based on observations made with the
IAC80 telescope operated on the island of Tenerife by the Instituto de
Astrofisica de Canarias in the Spanish Observatorio del Teide,
on observations obtained with the Apache Point Observatory 3.5-meter telescope,
which is owned and operated by the Astrophysical Research Consortium.
and on observations at Kitt Peak National Observatory, National Optical Astronomy
Observatory (NOAO Prop. IDs: 2011A-0392; PI: J. Coughlin), which is operated by the
Association of Universities for Research in Astronomy (AURA) under cooperative
agreement with the National Science Foundation.
IRAF is distributed by the National Optical Astronomy Observatory, which is operated
by the Association of Universities for Research in Astronomy (AURA) under a cooperative
agreement with the National Science Foundation. 
This research has made use of the SIMBAD database, operated at CDS, Strasbourg, France,
and of NASA's Astrophysics Data System Bibliographic Services.
Also, it used data products from the Two Micron All Sky Survey.
Some of the data presented in this paper were obtained from the Mikulski Archive
for Space Telescopes (MAST). This paper includes data collected by the Kepler mission.
Funding for the Kepler mission is provided by the NASA Science Mission directorate.
R.I.M. acknowledges support through the Programa de Acceso a Instalaciones Cientificas Singulares (E/309290).
J.L.C. acknowledges support through a National Science Foundation Graduate Research Fellowship.

\end{acknowledgements}

%
%

\bibliographystyle{aa} 
\bibliography{TCyg1} 

\begin{thebibliography}{95}
\expandafter\ifx\csname natexlab\endcsname\relax\def\natexlab#1{#1}\fi

\bibitem[{{Alonso} {et~al.}(2004){Alonso}, {Brown}, {Torres}, {Latham},
  {Sozzetti}, {Mandushev}, {Belmonte}, {Charbonneau}, {Deeg}, {Dunham},
  {O'Donovan}, \& {Stefanik}}]{alonso2004}
{Alonso}, R., {Brown}, T.~M., {Torres}, G., {et~al.} 2004, \apjl, 613, L153

\bibitem[{{Arribas} \& {Martinez Roger}(1988)}]{arribas1988}
{Arribas}, S. \& {Martinez Roger}, C. 1988, \aap, 206, 63

\bibitem[{{Ballesteros}(2012)}]{ballesteros2012}
{Ballesteros}, F.~J. 2012, EPL (Europhysics Letters), 97, 34008

\bibitem[{{Baraffe} {et~al.}(1998){Baraffe}, {Chabrier}, {Allard}, \&
  {Hauschildt}}]{baraffe1998}
{Baraffe}, I., {Chabrier}, G., {Allard}, F., \& {Hauschildt}, P.~H. 1998, \aap,
  337, 403

\bibitem[{{Bayless} \& {Orosz}(2006)}]{bayless2006}
{Bayless}, A.~J. \& {Orosz}, J.~A. 2006, \apj, 651, 1155

\bibitem[{{Behr}(2003)}]{behr2003}
{Behr}, B.~B. 2003, \apjs, 149, 67

\bibitem[{{Bessell}(1979)}]{bessell1979}
{Bessell}, M.~S. 1979, \pasp, 91, 589

\bibitem[{{Bessell}(1991)}]{bessell1991}
{Bessell}, M.~S. 1991, \aj, 101, 662

\bibitem[{{Birkby} {et~al.}(2012){Birkby}, {Nefs}, {Hodgkin}, {Kov{\'a}cs},
  {Sip{\H o}cz}, {Pinfield}, {Snellen}, {Mislis}, {Murgas}, {Lodieu}, {de
  Mooij}, {Goulding}, {Cruz}, {Stoev}, {Cappetta}, {Palle}, {Barrado},
  {Saglia}, {Martin}, \& {Pavlenko}}]{birkby2012}
{Birkby}, J., {Nefs}, B., {Hodgkin}, S., {et~al.} 2012, \mnras, 426, 1507

\bibitem[{Brogna(2006)}]{brogna2006}
Brogna, P. 2006, PhD thesis, San Francisco State University

\bibitem[{{Brown} {et~al.}(2011){Brown}, {Latham}, {Everett}, \&
  {Esquerdo}}]{brown2011}
{Brown}, T.~M., {Latham}, D.~W., {Everett}, M.~E., \& {Esquerdo}, G.~A. 2011,
  \aj, 142, 112

\bibitem[{{{\c C}ak{\i}rl{\i}}(2013)}]{cakirli2013b}
{{\c C}ak{\i}rl{\i}}, {\"O}. 2013, \na, 22, 15

\bibitem[{{{\c C}ak{\i}rl{\i}} {et~al.}(2013){{\c C}ak{\i}rl{\i}},
  {{\.I}bano{\v g}lu}, \& {Sipahi}}]{cakirli2013}
{{\c C}ak{\i}rl{\i}}, {\"O}., {{\.I}bano{\v g}lu}, C., \& {Sipahi}, E. 2013,
  \mnras, 429, 85

\bibitem[{{Carter} {et~al.}(2011){Carter}, {Fabrycky}, {Ragozzine}, {Holman},
  {Quinn}, {Latham}, {Buchhave}, {Van Cleve}, {Cochran}, {Cote}, {Endl},
  {Ford}, {Haas}, {Jenkins}, {Koch}, {Li}, {Lissauer}, {MacQueen}, {Middour},
  {Orosz}, {Rowe}, {Steffen}, \& {Welsh}}]{carter2011}
{Carter}, J.~A., {Fabrycky}, D.~C., {Ragozzine}, D., {et~al.} 2011, Science,
  331, 562

\bibitem[{{Claret} \& {Bloemen}(2011)}]{claret2011}
{Claret}, A. \& {Bloemen}, S. 2011, \aap, 529, A75

\bibitem[{{Clausen} {et~al.}(2009){Clausen}, {Bruntt}, {Claret}, {Larsen},
  {Andersen}, {Nordstr{\"o}m}, \& {Gim{\'e}nez}}]{clausen2009}
{Clausen}, J.~V., {Bruntt}, H., {Claret}, A., {et~al.} 2009, \aap, 502, 253

\bibitem[{{Coughlin}(2012)}]{coughlin2012}
{Coughlin}, J.~L. 2012, PhD thesis, New Mexico State University

\bibitem[{{Coughlin} {et~al.}(2011){Coughlin}, {L{\'o}pez-Morales}, {Harrison},
  {Ule}, \& {Hoffman}}]{coughlin2011}
{Coughlin}, J.~L., {L{\'o}pez-Morales}, M., {Harrison}, T.~E., {Ule}, N., \&
  {Hoffman}, D.~I. 2011, \aj, 141, 78

\bibitem[{{Cox}(2000)}]{cox2000}
{Cox}, A.~N. 2000, {Allen's Astrophysical Quantities}, 4th edn. (New York: AIP)

\bibitem[{Cutri(2003)}]{cutri2003}
Cutri, R.M., e.~a. 2003, \emph{The IRSA 2MASS All-Sky Point Source Catalogue},
  Tech. rep., NASA/IPAC Infrared Science Archive, available at
  \url{http://irsa.ipac.caltech.edu/applications/Gator/}

\bibitem[{{Devor}(2008)}]{devor2008thesis}
{Devor}, J. 2008, PhD thesis, Harvard University

\bibitem[{{Devor} {et~al.}(2008{\natexlab{a}}){Devor}, {Charbonneau},
  {O'Donovan}, {Mandushev}, \& {Torres}}]{devor2008}
{Devor}, J., {Charbonneau}, D., {O'Donovan}, F.~T., {Mandushev}, G., \&
  {Torres}, G. 2008{\natexlab{a}}, \aj, 135, 850

\bibitem[{{Devor} {et~al.}(2008{\natexlab{b}}){Devor}, {Charbonneau}, {Torres},
  {Blake}, {White}, {Rabus}, {O'Donovan}, {Mandushev}, {Bakos}, {F{\H
  u}r{\'e}sz}, \& {Szentgyorgyi}}]{devor2008b}
{Devor}, J., {Charbonneau}, D., {Torres}, G., {et~al.} 2008{\natexlab{b}},
  \apj, 687, 1253

\bibitem[{{Diaz-Cordoves} \& {Gimenez}(1992)}]{diaz-cordoves1992}
{Diaz-Cordoves}, J. \& {Gimenez}, A. 1992, \aap, 259, 227

\bibitem[{{Dotter} {et~al.}(2008){Dotter}, {Chaboyer}, {Jevremovi{\'c}},
  {Kostov}, {Baron}, \& {Ferguson}}]{dotter2008}
{Dotter}, A., {Chaboyer}, B., {Jevremovi{\'c}}, D., {et~al.} 2008, \apjs, 178,
  89

\bibitem[{{Eastman} {et~al.}(2010){Eastman}, {Siverd}, \&
  {Gaudi}}]{eastman2010}
{Eastman}, J., {Siverd}, R., \& {Gaudi}, B.~S. 2010, \pasp, 122, 935

\bibitem[{{Eggen}(1989)}]{eggen1989}
{Eggen}, O.~J. 1989, \pasp, 101, 366

\bibitem[{{Eggleton}(1983)}]{eggleton1983}
{Eggleton}, P.~P. 1983, \apj, 268, 368

\bibitem[{{Eisenstein} {et~al.}(2011){Eisenstein}, {Weinberg}, {Agol},
  {Aihara}, {Allende Prieto}, {Anderson}, {Arns}, {Aubourg}, {Bailey},
  {Balbinot}, \& et~al.}]{eisenstein2011}
{Eisenstein}, D.~J., {Weinberg}, D.~H., {Agol}, E., {et~al.} 2011, \aj, 142, 72

\bibitem[{{Etzel}(1981)}]{etzel1981}
{Etzel}, P.~B. 1981, in Photometric and Spectroscopic Binary Systems, ed. E.~B.
  {Carling} \& Z.~{Kopal}, 111

\bibitem[{{Everett} {et~al.}(2012){Everett}, {Howell}, \&
  {Kinemuchi}}]{everett2012}
{Everett}, M.~E., {Howell}, S.~B., \& {Kinemuchi}, K. 2012, \pasp, 124, 316

\bibitem[{{Flower}(1996)}]{flower1996}
{Flower}, P.~J. 1996, \apj, 469, 355

\bibitem[{{Fukugita} {et~al.}(2011){Fukugita}, {Yasuda}, {Doi}, {Gunn}, \&
  {York}}]{fukugita2011}
{Fukugita}, M., {Yasuda}, N., {Doi}, M., {Gunn}, J.~E., \& {York}, D.~G. 2011,
  \aj, 141, 47

\bibitem[{{Gelman} \& {Rubin}(1992)}]{gelman1992}
{Gelman}, A. \& {Rubin}, D. 1992, Statistical Science, 7, 457

\bibitem[{{Girardi} {et~al.}(2000){Girardi}, {Bressan}, {Bertelli}, \&
  {Chiosi}}]{girardi2000}
{Girardi}, L., {Bressan}, A., {Bertelli}, G., \& {Chiosi}, C. 2000, \aaps, 141,
  371

\bibitem[{{G{\'o}mez Maqueo Chew} {et~al.}(2014){G{\'o}mez Maqueo Chew},
  {Morales}, {Faedi}, {Garc{\'{\i}}a-Melendo}, {Hebb}, {Rodler}, {Deshpande},
  {Mahadevan}, {McCormac}, {Barnes}, {Triaud}, {Lopez-Morales}, {Skillen},
  {Collier Cameron}, {Joner}, {Laney}, {Stephens}, {Stassun}, {Cargile}, \&
  {Monta{\~n}{\'e}s-Rodr{\'{\i}}guez}}]{gomez2014}
{G{\'o}mez Maqueo Chew}, Y., {Morales}, J.~C., {Faedi}, F., {et~al.} 2014,
  \aap, 572, A50

\bibitem[{{Granzer} {et~al.}(2000){Granzer}, {Sch{\"u}ssler}, {Caligari}, \&
  {Strassmeier}}]{granzer2000}
{Granzer}, T., {Sch{\"u}ssler}, M., {Caligari}, P., \& {Strassmeier}, K.~G.
  2000, \aap, 355, 1087

\bibitem[{{Hatzes}(1995)}]{hatzes1995}
{Hatzes}, A.~P. 1995, in IAU Symposium, Vol. 176, IAU Symposium, 90P

\bibitem[{{He{\l}miniak} {et~al.}(2014){He{\l}miniak}, {Brahm}, {Ratajczak},
  {Espinoza}, {Jord{\'a}n}, {Konacki}, \& {Rabus}}]{helminiak2014}
{He{\l}miniak}, K.~G., {Brahm}, R., {Ratajczak}, M., {et~al.} 2014, \aap, 567,
  A64

\bibitem[{{He{\l}miniak} \& {Konacki}(2011)}]{helminiak2011}
{He{\l}miniak}, K.~G. \& {Konacki}, M. 2011, \aap, 526, A29

\bibitem[{{He{\l}miniak} {et~al.}(2012){He{\l}miniak}, {Konacki},
  {R{\'o}{\.Z}yczka}, {Ka{\l}u{\.Z}ny}, {Ratajczak}, {Borkowski}, {Sybilski},
  {Muterspaugh}, {Reichart}, {Ivarsen}, {Haislip}, {Crain}, {Foster},
  {Nysewander}, \& {LaCluyze}}]{helminiak2012}
{He{\l}miniak}, K.~G., {Konacki}, M., {R{\'o}{\.Z}yczka}, M., {et~al.} 2012,
  \mnras, 425, 1245

\bibitem[{{He{\l}miniak} {et~al.}(2011){He{\l}miniak}, {Konacki},
  {Z{\l}oczewski}, {Ratajczak}, {Reichart}, {Ivarsen}, {Haislip}, {Crain},
  {Foster}, {Nysewander}, \& {Lacluyze}}]{helminiak2011b}
{He{\l}miniak}, K.~G., {Konacki}, M., {Z{\l}oczewski}, K., {et~al.} 2011, \aap,
  527, A14

\bibitem[{{Houdashelt} {et~al.}(2000){Houdashelt}, {Bell}, \&
  {Sweigart}}]{houdashelt2000}
{Houdashelt}, M.~L., {Bell}, R.~A., \& {Sweigart}, A.~V. 2000, \aj, 119, 1448

\bibitem[{{Hut}(1981)}]{hut1981}
{Hut}, P. 1981, \aap, 99, 126

\bibitem[{{IAU Inter-Division A-G Working Group on Nominal Units for Stellar \&
  Planetary Astronomy}(2015)}]{IAU_B2_2015}
{IAU Inter-Division A-G Working Group on Nominal Units for Stellar \& Planetary
  Astronomy}. 2015, {Resolution B2 on recommended zero points for the absolute
  and apparent bolometric magnitude scales}, Tech. rep., IAU

\bibitem[{{Iglesias-Marzoa} {et~al.}(2015){Iglesias-Marzoa},
  {L{\'o}pez-Morales}, \& {Jes{\'u}s Ar{\'e}valo
  Morales}}]{iglesias-marzoa2015}
{Iglesias-Marzoa}, R., {L{\'o}pez-Morales}, M., \& {Jes{\'u}s Ar{\'e}valo
  Morales}, M. 2015, \pasp, 127, 567

\bibitem[{{Irwin} {et~al.}(2009){Irwin}, {Charbonneau}, {Berta}, {Quinn},
  {Latham}, {Torres}, {Blake}, {Burke}, {Esquerdo}, {F{\"u}r{\'e}sz}, {Mink},
  {Nutzman}, {Szentgyorgyi}, {Calkins}, {Falco}, {Bloom}, \&
  {Starr}}]{irwin2009}
{Irwin}, J., {Charbonneau}, D., {Berta}, Z.~K., {et~al.} 2009, \apj, 701, 1436

\bibitem[{{Irwin} {et~al.}(2011){Irwin}, {Quinn}, {Berta}, {Latham}, {Torres},
  {Burke}, {Charbonneau}, {Dittmann}, {Esquerdo}, {Stefanik}, {Oksanen},
  {Buchhave}, {Nutzman}, {Berlind}, {Calkins}, \& {Falco}}]{irwin2011}
{Irwin}, J.~M., {Quinn}, S.~N., {Berta}, Z.~K., {et~al.} 2011, \apj, 742, 123

\bibitem[{{Ivezi{\'c}} {et~al.}(2008){Ivezi{\'c}}, {Sesar}, {Juri{\'c}},
  {Bond}, {Dalcanton}, {Rockosi}, {Yanny}, {Newberg}, {Beers}, {Allende
  Prieto}, {Wilhelm}, {Lee}, {Sivarani}, {Norris}, {Bailer-Jones}, {Re
  Fiorentin}, {Schlegel}, {Uomoto}, {Lupton}, {Knapp}, {Gunn}, {Covey},
  {Smith}, {Miknaitis}, {Doi}, {Tanaka}, {Fukugita}, {Kent}, {Finkbeiner},
  {Munn}, {Pier}, {Quinn}, {Hawley}, {Anderson}, {Kiuchi}, {Chen}, {Bushong},
  {Sohi}, {Haggard}, {Kimball}, {Barentine}, {Brewington}, {Harvanek},
  {Kleinman}, {Krzesinski}, {Long}, {Nitta}, {Snedden}, {Lee}, {Harris},
  {Brinkmann}, {Schneider}, \& {York}}]{ivezic2008}
{Ivezi{\'c}}, {\v Z}., {Sesar}, B., {Juri{\'c}}, M., {et~al.} 2008, \apj, 684,
  287

\bibitem[{{Jenkins} {et~al.}(2010){Jenkins}, {Caldwell}, {Chandrasekaran},
  {Twicken}, {Bryson}, {Quintana}, {Clarke}, {Li}, {Allen}, {Tenenbaum}, {Wu},
  {Klaus}, {Middour}, {Cote}, {McCauliff}, {Girouard}, {Gunter}, {Wohler},
  {Sommers}, {Hall}, {Uddin}, {Wu}, {Bhavsar}, {Van Cleve}, {Pletcher},
  {Dotson}, {Haas}, {Gilliland}, {Koch}, \& {Borucki}}]{jenkins2010}
{Jenkins}, J.~M., {Caldwell}, D.~A., {Chandrasekaran}, H., {et~al.} 2010,
  \apjl, 713, L87

\bibitem[{{Johnson} \& {Soderblom}(1987)}]{johnson1987}
{Johnson}, D.~R.~H. \& {Soderblom}, D.~R. 1987, \aj, 93, 864

\bibitem[{{Kipping}(2010)}]{kipping2010}
{Kipping}, D.~M. 2010, \mnras, 408, 1758

\bibitem[{{Kraus} {et~al.}(2011){Kraus}, {Tucker}, {Thompson}, {Craine}, \&
  {Hillenbrand}}]{kraus2011}
{Kraus}, A.~L., {Tucker}, R.~A., {Thompson}, M.~I., {Craine}, E.~R., \&
  {Hillenbrand}, L.~A. 2011, \apj, 728, 48

\bibitem[{Kurucz(1993)}]{kurucz1993}
Kurucz, R.~L. 1993, \emph{The 1993 Kurucz Stellar Atmospheres Atlas}, available
  at \url{ftp://ftp.stsci.edu/cdbs/grid/kurucz93models}

\bibitem[{{Landolt}(2009)}]{landolt2009}
{Landolt}, A.~U. 2009, \aj, 137, 4186

\bibitem[{{Lanza} {et~al.}(2014){Lanza}, {Das Chagas}, \& {De
  Medeiros}}]{lanza2014}
{Lanza}, A.~F., {Das Chagas}, M.~L., \& {De Medeiros}, J.~R. 2014, \aap, 564,
  A50

\bibitem[{{Leggett}(1992)}]{leggett1992}
{Leggett}, S.~K. 1992, \apjs, 82, 351

\bibitem[{{Lejeune} {et~al.}(1998){Lejeune}, {Cuisinier}, \&
  {Buser}}]{lejeune1998}
{Lejeune}, T., {Cuisinier}, F., \& {Buser}, R. 1998, \aaps, 130, 65

\bibitem[{{L{\'o}pez-Morales} \& {Bonanos}(2008)}]{lopez-morales2008}
{L{\'o}pez-Morales}, M. \& {Bonanos}, A.~Z. 2008, \apjl, 685, L47

\bibitem[{{L{\'o}pez-Morales} \& {Ribas}(2005)}]{lopez-morales2005}
{L{\'o}pez-Morales}, M. \& {Ribas}, I. 2005, \apj, 631, 1120

\bibitem[{{Maldonado} {et~al.}(2010){Maldonado}, {Mart{\'{\i}}nez-Arn{\'a}iz},
  {Eiroa}, {Montes}, \& {Montesinos}}]{maldonado2010}
{Maldonado}, J., {Mart{\'{\i}}nez-Arn{\'a}iz}, R.~M., {Eiroa}, C., {Montes},
  D., \& {Montesinos}, B. 2010, \aap, 521, A12

\bibitem[{{Mamajek}(2015)}]{mamajek2015}
{Mamajek}, E.~E. 2015, \emph{A Modern Mean Stellar Color and Effective
  Temperature Sequence for O9V-Y0V Dwarf Stars}, available at
  \url{http://www.pas.rochester.edu/~emamajek/EEM_dwarf_UBVIJHK_colors_Teff.tx%
t}

\bibitem[{{Masana} {et~al.}(2006){Masana}, {Jordi}, \& {Ribas}}]{masana2006}
{Masana}, E., {Jordi}, C., \& {Ribas}, I. 2006, \aap, 450, 735

\bibitem[{{Merline} \& {Howell}(1995)}]{merline1995}
{Merline}, W.~J. \& {Howell}, S.~B. 1995, Experimental Astronomy, 6, 163

\bibitem[{{Montes} {et~al.}(2001){Montes}, {L{\'o}pez-Santiago}, {G{\'a}lvez},
  {Fern{\'a}ndez-Figueroa}, {De Castro}, \& {Cornide}}]{montes2001}
{Montes}, D., {L{\'o}pez-Santiago}, J., {G{\'a}lvez}, M.~C., {et~al.} 2001,
  \mnras, 328, 45

\bibitem[{{Morales} {et~al.}(2010){Morales}, {Gallardo}, {Ribas}, {Jordi},
  {Baraffe}, \& {Chabrier}}]{morales2010}
{Morales}, J.~C., {Gallardo}, J., {Ribas}, I., {et~al.} 2010, \apj, 718, 502

\bibitem[{{Morales} {et~al.}(2009{\natexlab{a}}){Morales}, {Ribas}, {Jordi},
  {Torres}, {Gallardo}, {Guinan}, {Charbonneau}, {Wolf}, {Latham},
  {Anglada-Escud{\'e}}, {Bradstreet}, {Everett}, {O'Donovan}, {Mandushev}, \&
  {Mathieu}}]{morales2009}
{Morales}, J.~C., {Ribas}, I., {Jordi}, C., {et~al.} 2009{\natexlab{a}}, \apj,
  691, 1400

\bibitem[{{Morales} {et~al.}(2009{\natexlab{b}}){Morales}, {Torres},
  {Marschall}, \& {Brehm}}]{morales2009b}
{Morales}, J.~C., {Torres}, G., {Marschall}, L.~A., \& {Brehm}, W.
  2009{\natexlab{b}}, \apj, 707, 671

\bibitem[{{Moro} \& {Munari}(2000)}]{moro2000}
{Moro}, D. \& {Munari}, U. 2000, \aaps, 147, 361

\bibitem[{{Munari} {et~al.}(2005){Munari}, {Sordo}, {Castelli}, \&
  {Zwitter}}]{munari2005}
{Munari}, U., {Sordo}, R., {Castelli}, F., \& {Zwitter}, T. 2005, \aap, 442,
  1127

\bibitem[{{Popper} \& {Etzel}(1981)}]{popper1981}
{Popper}, D.~M. \& {Etzel}, P.~B. 1981, \aj, 86, 102

\bibitem[{{Pr{\v s}a} {et~al.}(2011){Pr{\v s}a}, {Batalha}, {Slawson}, {Doyle},
  {Welsh}, {Orosz}, {Seager}, {Rucker}, {Mjaseth}, {Engle}, {Conroy},
  {Jenkins}, {Caldwell}, {Koch}, \& {Borucki}}]{prsa2011b}
{Pr{\v s}a}, A., {Batalha}, N., {Slawson}, R.~W., {et~al.} 2011, \aj, 141, 83

\bibitem[{{Pr{\v s}a} {et~al.}(2008){Pr{\v s}a}, {Guinan}, {Devinney},
  {DeGeorge}, {Bradstreet}, {Giammarco}, {Alcock}, \& {Engle}}]{prsa2008}
{Pr{\v s}a}, A., {Guinan}, E.~F., {Devinney}, E.~J., {et~al.} 2008, \apj, 687,
  542

\bibitem[{{Pr{\v s}a} {et~al.}(2016){Pr{\v s}a}, {Harmanec}, {Torres},
  {Mamajek}, {Asplund}, {Capitaine}, {Christensen-Dalsgaard}, {Depagne},
  {Haberreiter}, {Hekker}, {Hilton}, {Kopp}, {Kostov}, {Kurtz}, {Laskar},
  {Mason}, {Milone}, {Montgomery}, {Richards}, {Schmutz}, {Schou}, \&
  {Stewart}}]{IAU_B3_2015}
{Pr{\v s}a}, A., {Harmanec}, P., {Torres}, G., {et~al.} 2016, \aj, 152, 41

\bibitem[{{Pr{\v s}a} \& {Zwitter}(2005)}]{prsa2005}
{Pr{\v s}a}, A. \& {Zwitter}, T. 2005, \apj, 628, 426

\bibitem[{Pr\v{s}a(2011)}]{prsa2011}
Pr\v{s}a, A. 2011, \emph{PHOEBE Scientific Reference}

\bibitem[{{Reinhold} \& {Arlt}(2015)}]{reinhold2015}
{Reinhold}, T. \& {Arlt}, R. 2015, \aap, 576, A15

\bibitem[{{Ribas}(2003)}]{ribas2003}
{Ribas}, I. 2003, \aap, 398, 239

\bibitem[{{Rozyczka} {et~al.}(2009){Rozyczka}, {Kaluzny}, {Pietrukowicz},
  {Pych}, {Mazur}, {Catelan}, \& {Thompson}}]{rozyczka2009}
{Rozyczka}, M., {Kaluzny}, J., {Pietrukowicz}, P., {et~al.} 2009, \actaa, 59,
  385

\bibitem[{{Ruci{\'n}ski}(1969)}]{rucinski1969b}
{Ruci{\'n}ski}, S.~M. 1969, \actaa, 19, 245

\bibitem[{{Scargle}(1982)}]{scargle1982}
{Scargle}, J.~D. 1982, \apj, 263, 835

\bibitem[{{Skrutskie} {et~al.}(2006){Skrutskie}, {Cutri}, {Stiening},
  {Weinberg}, {Schneider}, {Carpenter}, {Beichman}, {Capps}, {Chester},
  {Elias}, {Huchra}, {Liebert}, {Lonsdale}, {Monet}, {Price}, {Seitzer},
  {Jarrett}, {Kirkpatrick}, {Gizis}, {Howard}, {Evans}, {Fowler}, {Fullmer},
  {Hurt}, {Light}, {Kopan}, {Marsh}, {McCallon}, {Tam}, {Van Dyk}, \&
  {Wheelock}}]{skrutskie2006}
{Skrutskie}, M.~F., {Cutri}, R.~M., {Stiening}, R., {et~al.} 2006, \aj, 131,
  1163

\bibitem[{{Skuljan} {et~al.}(1999){Skuljan}, {Hearnshaw}, \&
  {Cottrell}}]{skuljan1999}
{Skuljan}, J., {Hearnshaw}, J.~B., \& {Cottrell}, P.~L. 1999, \mnras, 308, 731

\bibitem[{{Slawson} {et~al.}(2011){Slawson}, {Pr{\v s}a}, {Welsh}, {Orosz},
  {Rucker}, {Batalha}, {Doyle}, {Engle}, {Conroy}, {Coughlin}, {Gregg},
  {Fetherolf}, {Short}, {Windmiller}, {Fabrycky}, {Howell}, {Jenkins}, {Uddin},
  {Mullally}, {Seader}, {Thompson}, {Sanderfer}, {Borucki}, \&
  {Koch}}]{slawson2011}
{Slawson}, R.~W., {Pr{\v s}a}, A., {Welsh}, W.~F., {et~al.} 2011, \aj, 142, 160

\bibitem[{{Southworth} {et~al.}(2004){Southworth}, {Maxted}, \&
  {Smalley}}]{southworth2004}
{Southworth}, J., {Maxted}, P.~F.~L., \& {Smalley}, B. 2004, \mnras, 351, 1277

\bibitem[{{Torres}(2010)}]{torres2010}
{Torres}, G. 2010, \aj, 140, 1158

\bibitem[{{Torres} \& {Ribas}(2002)}]{torres2002}
{Torres}, G. \& {Ribas}, I. 2002, \apj, 567, 1140

\bibitem[{{Vaccaro} {et~al.}(2007){Vaccaro}, {Rudkin}, {Kawka}, {Vennes},
  {Oswalt}, {Silver}, {Wood}, \& {Smith}}]{vaccaro2007}
{Vaccaro}, T.~R., {Rudkin}, M., {Kawka}, A., {et~al.} 2007, \apj, 661, 1112

\bibitem[{{van Dokkum}(2001)}]{vandokkum2001}
{van Dokkum}, P.~G. 2001, \pasp, 113, 1420

\bibitem[{{van Hamme}(1993)}]{vanhamme1993}
{van Hamme}, W. 1993, \aj, 106, 2096

\bibitem[{{Yi} {et~al.}(2001){Yi}, {Demarque}, {Kim}, {Lee}, {Ree}, {Lejeune},
  \& {Barnes}}]{yi2001}
{Yi}, S., {Demarque}, P., {Kim}, Y.-C., {et~al.} 2001, \apjs, 136, 417

\bibitem[{{Zahn}(1977)}]{zahn1977}
{Zahn}, J.-P. 1977, \aap, 57, 383

\bibitem[{{Zucker} \& {Mazeh}(1994)}]{zucker1994}
{Zucker}, S. \& {Mazeh}, T. 1994, \apj, 420, 806

\bibitem[{{Östensen}(2000)}]{ostensen2000thesis}
{Östensen}, R. 2000, PhD thesis, University of Trömso

\bibitem[{{Östensen} \& {Solheim}(2000)}]{ostensen2000}
{Östensen}, R. \& {Solheim}, J.-E. 2000, Baltic Astronomy, 9, 411

\end{thebibliography}

\end{document}